\documentclass[a4paper,11pt]{article}
\usepackage{jheppub}
\usepackage{adjustbox, slashed, siunitx}
\usepackage{booktabs, makecell}
\usepackage{cleveref}
\usepackage{tikz}
\usetikzlibrary{positioning, calc}

\makeatletter\g@addto@macro\bfseries{\boldmath}\makeatother

\newcommand{\myr}[1]{{\color{red!50!black}#1}}

\newcommand{\myg}[1]{{\color{green!50!black}#1}}
\usepackage{ifpdf,feynmp}
\ifpdf\fi
\AtBeginDocument{\begin{fmffile}{fgraph}
\fmfcmd{%
arrow_ang := 11;
arrow_len := 3.5thick;
curly_len := 2.5thick;
color myg,myp,myr;
myg := .5green;
myp := .7blue;
myr := .5red;
style_def top expr p =
  save oldpen;
  pen oldpen;
  oldpen := currentpen;
  pickup oldpen scaled 1.25;
  ccutdraw p;
  pickup oldpen scaled .25;
  cullit; undraw p; cullit;
  cfill (arrow p);
  pickup oldpen scaled 1;
enddef;
}}

\def\sectionautorefname~#1\null{sec.\:#1\null}
\def\subsectionautorefname~#1\null{sec.\:#1\null}
\def\figureautorefname~#1\null{fig.\:#1\null}
\def\tableautorefname~#1\null{table\:#1\null}
\def\equationautorefname~#1\null{eq.\:(#1)\null}

\newcommand{\minitab}[2][c]{\begin{tabular}{@{}#1@{}}#2\end{tabular}}

\newcommand{\LOQCD}{\text{LO QCD}}
\newcommand{\NLOQCD}{\text{NLO QCD}}

\newcommand{\ttop}{\ensuremath{\bar tt\bar tW^+}}

\title{%
The inseparable three and four tops
}
\author[a]{Gauthier Durieux,}
\author[b]{Hesham El Faham,}
\author[c]{Rikkert Frederix,}
\author[d]{Davide Pagani,}
\author[b]{Marco Zaro}

\affiliation[a]{Centre for Cosmology, Particle Physics and Phenomenology, Université catholique de Louvain,\\ 1348 Louvain-la-Neuve, Belgium}
\affiliation[b]{TIFLab, Universit\`a degli Studi di Milano and INFN, Sezione di Milano,\\ Via Celoria 16, 20133 Milano, Italy}
\affiliation[c]{Division of Particle and Nuclear Physics, Department of Physics, Lund University,\\ Box 118, SE-221 00 Lund, Sweden}
\affiliation[d]{INFN, Sezione di Bologna,\\ Via Irnerio 46, 40126 Bologna, Italy}

\preprint{%
\begin{flushright}%
IRMP-CP3-26-20\\
TIF-UNIMI-2026-27
\end{flushright}%
}%

\abstract{%
In measurements of four-top-quark production ($tttt$), LHC collaborations observe a significant degeneracy with three-top-quark production.
We compute the dominant three-top-production mode, namely associated production with a $W$ boson ($tttW$), at complete next-to-leading order (NLO), including all possible QCD and electroweak (EW) corrections.
Beyond leading order (LO), $tttW$ production with the radiation of an additional $b$-flavoured quark contributes to the same final state as $tttt$ production with a $t \to bW$ decay.
Away from the on-shell top-quark limit, the usual overlap removal of resonant contributions in the non-resonant computation either breaks gauge invariance and generates unitarity violation or involves a significant arbitrariness in the required reshuffling of momenta.
To overcome these issues, we introduce a novel window-removal prescription that produces consistent predictions for the inseparable $tttW+tttt$ process, with both components described at NLO accuracy.
We argue that such a joint prediction should be used in comparisons with experimental selections targeting $tttt$ production, since the on-shell $tttt$ component can not be isolated in practice.
Such a joint prediction has an inclusive rate more than $10\%$ higher than the purely on-shell $tttt$ one.
We also study an idealised veto on additional hard and central $b$-jet radiation, which suppresses the contributions of resonant $tttt$ diagrams as well as their interference with non-resonant $tttW$ ones and therefore defines a relatively pure $tttW$-like signal region.
Formally subleading coupling orders are numerically important at LO, while the corresponding subleading NLO corrections largely cancel both inclusively and differentially.
Consequently, the complete-NLO prediction is well approximated by retaining the first three LO coupling orders together with the leading QCD NLO correction.
}%

\begin{document}
\maketitle
\flushbottom

\section{Introduction}
The top quark occupies a special position at the upper edge of the known elementary particle spectrum and may have enhanced couplings to physics not yet accounted for in the Standard Model (SM).
Its study therefore constitutes one of the main pillars of the physics programme at the Large Hadron Collider (LHC) where it is primarily produced in pairs and singly.
The pair production mechanism is dominantly due to the strong interaction, while single production is a weak process.

The production of four top quarks is instead much rarer, with a SM cross section five orders of magnitude smaller than that of the pair production mode.
However, this rare process has distinctive features.
For example, it constrains the top-quark Yukawa coupling with the Higgs boson, lifting the dependence on the Higgs decay width~\cite{Cao:2016wib, Cao:2019ygh, ATLAS:2024mhs} and makes it possible to directly access beyond-the-SM (BSM) effects that are only indirectly probed by the dominant production mechanisms (i.e.\ via loop effects)~\cite{Zhang:2017mls, Malekhosseini:2018fgp, Darme:2018dvz, Hou:2019gpn, Cao:2019qrb, Alvarez:2019uxp, Banelli:2020iau, Hou:2020chc, Darme:2021gtt, Cao:2021qqt, Carpenter:2021vga, Aoude:2022deh, Bally:2022naz, Aleshko:2023rkv, Choudhury:2024mox, Degrande:2024mbg, DiNoi:2025uhu, Darme:2025leu}.
For a review on four-top production, see also~\cite{Blekman:2022jag}.

Four-top production has been computed in the SM at next-to-leading order (NLO) both in the strong and electroweak couplings~\cite{Bevilacqua:2012em, Maltoni:2015ena, Frederix:2017wme}, with threshold resummation~\cite{vanBeekveld:2022hty, vanBeekveld:2025ghw}, and top-quark decays described at NLO QCD in the narrow width approximation~\cite{Alsairafi:2025rjd}.
ATLAS and CMS reported its observation in 2023~\cite{ATLAS:2023ajo,CMS:2023ftu}.
This task was made particularly challenging by the small rate, large number of decay products, and significant probability that some of these are lost outside of the detector acceptance or misidentified.
Given the complexity of the final-state at hand, these analyses heavily relied on machine-learning techniques to enhance selection efficiencies and separate the signal from backgrounds.

Three-top-quark production is an even rarer process in the SM, with a cross section about a factor of five smaller than that of four-top production.
It was therefore not considered a major background in the design of four-top experimental analyses.
However, it turned out that the kinematics of these two processes are very similar.
This made it extremely challenging to disentangle the two contributions, resulting in a significant degeneracy between their extracted signal strengths~\cite{ATLAS:2023ajo, CMS:2023ftu, CMS:2025rug}.
They were practically inseparable.
The observation of four-top production could therefore only be claimed by fixing the three-top production cross section to its SM value.
Given that three-top production could only be modelled at LO, uncertainties still had a sizeable impact on the obtained four-top significance.
Incidentally, a small excess of signal events over SM expectations is present in ATLAS data~\cite{ATLAS:2023ajo}, which also mildly prefers an enhanced three-top rate (by more than an order of magnitude) and a vanishing four-top one.

BSM physics may also intertwine three and four top-quark productions.
Their simultaneous modifications (together with $t\bar{t}b\bar{b}$ production too) can be expected in BSM models where the left-handed top quark couples to new states, since it forms a $\text{SU}(2)_L$ doublet with the left-handed bottom quark.
The simplest examples are models featuring a new scalar coupling to a top-quark pair which, by Lorentz invariance, involves both chiralities (see e.g.\ the recent analysis in~\cite{Darme:2025leu}).
The $\bar{Q}_L t_R \Phi$ coupling where $Q_L\equiv(t_L, b_L)$ and $\Phi\equiv(\Phi^0,\Phi^-)$ then contains both $\bar{t}_Lt_R \Phi^0$ and $\bar{b}_L t_R \Phi^-$ interactions.

Beyond leading order (LO), the inseparability between three- and four-top production acquires a more fundamental character.
Radiating an extra $b$ quark in the leading $tttW$ three-top production channel\footnote{For simplicity, we only label differently particles and anti-particles when necessary: e.g.\ $t\bar{t}tW^{-}$ and $\bar{t}t\bar{t}W^{+}$ are referred to jointly as $tttW$.} (in the five-flavour scheme) indeed produces the exact same $tttWb$ final state as four-top $tttt$ production followed by a $t\to Wb$ decay.
The resonant $tttt$ and non-resonant $tttW$ diagrams then constitute genuinely inseparable families contributing to the same amplitude.
Such NLO overlaps between processes appearing distinct at LO is well-known to arise in simpler final states such as the $tWb$ one, which receives contributions from both $tW$ with an extra $b$ radiation and top-quark pair production followed by a $t\to Wb$ decay.

An obvious possibility to consistently include all contributions simultaneously is to define the final state in terms of  $W$'s and $b$'s only (in the four-flavour scheme).
Although the NLO calculation of $WbWb$ production is manageable, the $WbWbWbWb$ one would be hard.
Instead, it is more efficient to artificially split the prediction in components each separately computed at NLO, before putting them together.
The resonant $tttt$ production constitutes a first well-defined component in the strict on-shell limit.
To form the second component, the overlap between the NLO computation of $tttW$ and this first on-shell $tttt$ component has to be removed.
Various overlap removal procedures exist but they either involve much arbitrariness or lead to sizeable unitarity violation, unlike in previous studies of $tWb+X$ processes~\cite{Frixione:2008yi, White:2009yt, Re:2010bp, Hollik:2012rc, Frixione:2019fxg, Faham:2021zet}.
We therefore propose a new prescription which is both simple and well-behaved.
It allows to produce joint $tttW+tttt$ predictions that are NLO accurate both in the on-shell and off-shell regime.

Although it is inseparable from $tttt$ production in general kinematics, the $tttW$ process could acquire an effective physical meaning in a phase-space restriction where resonant $tttt$ diagrams and their interference with non-resonant $tttW$ ones are much suppressed.
A measurement in such a region would provide a more granular test of the SM than the inclusive measurement of the joint $tttW+tttt$ rate.
To attempt such a restricted phase-space definition, we take inspiration from the simpler cases of $tW+X$ processes, where a veto on hard and central $b$-jet radiation proved effective.
In $tttWb$ however, top-quark decays already generate three $b$ jets.
Reliably identifying and vetoing the fourth radiated one is thus most probably impractical.
Nevertheless, an idealised $b$ veto is straightforward to impose on simulation.
As we will see, it indeed effectively suppresses resonant $tttt$ diagrams and their interference with non-resonant $tttW$ ones.
It could thus be used as target for the definition of a realistic $tttW$-enriched selection, possibly using machine-learning techniques.
If it is not sufficiently pure, data in that phase-space region should still be compared to the joint $tttW+tttt$ prediction.

With these consistent prescriptions at hand, we will eventually be able to address the computation of NLO predictions both in QCD and electroweak (EW) couplings.
We first examine the factorisation and renormalisation scale choice at NLO in QCD, before exploring the complete set of LO and NLO coupling orders.
As in four-top production~\cite{Frederix:2017wme}, a rather intricate pattern of NLO corrections is found.
Several formally subleading EW orders are actually sizeable, due to the large Yukawa coupling of the top quark.
Contributions at different coupling orders also appear with opposite signs, sometimes causing high levels of cancellation.

The rest of this paper is organised as follows: in Sec.~\ref{sec:process_def} we introduce the three-top final state and describe its features, focusing on the overlap with four-top arising at NLO, and on the prescriptions to handle it.
Results of the complete NLO calculation are presented in Sec.~\ref{sec:complete-nlo}, first discussing the scale choice, then examining integrated and differential rates at complete-NLO accuracy.
Conclusions are drawn in Sec.~\ref{sec:concl}.

\section{Process definitions}
\label{sec:process_def}
The three-top-quark production process can only be defined unambiguously at LO and in the five-flavour scheme (5FS) where the bottom quark is considered massless, treated on the same footing as lighter quarks, and included in the parton distribution function (PDF) of the proton.
In the four-flavour scheme, the three-top production in association with a $W$ boson always features a final-state $b$ quark ($tttWb$) and already overlaps at LO with $tttt$.
The 5FS is adopted throughout this paper.

\begin{figure}[tb]\centering
\fmfframe(5,5)(5,5){\begin{fmfgraph*}(25,15)
\fmfleft{l1,l2}
\fmfright{r1,r2,r3,r4}
\fmf{fermion,tens=2}{l1,v1,l2}\fmflabel{$u$}{l1}\fmflabel{$\bar{d}$}{l2}
\fmf{photon, lab=$W$, tens=4}{v1,v2}
\fmf{top, fore=myr, lab=$ $, l.side=right, tens=2}{v2,v3}
\fmf{top, fore=myr}{v3,r1}\fmflabel{$\myr{t}$}{r1}
\fmf{fermion, fore=myg}{r4,v2}\fmflabel{$\myg{\bar{b}}$}{r4}
\fmffreeze
\fmf{gluon, tens=2}{v3,v4}
\fmf{top, fore=myr}{r2,v4,r3}\fmflabel{$\myr{\bar{t}}$}{r2}\fmflabel{$\myr{t}$}{r3}
\end{fmfgraph*}}
\fmfframe(5,5)(5,5){\begin{fmfgraph*}(25,15)
\fmfleft{l1,l2}
\fmfright{r1,r2,r3,r4}
\fmf{fermion,tens=2, fore=myg}{l1,v1}\fmflabel{$\myg{b}$}{l1}
\fmf{fermion,tens=2}{l2,v2}\fmflabel{$u$}{l2}
\fmf{fermion}{v8,r4}\fmf{vanilla}{v2,v8}\fmflabel{$d$}{r4}
\fmf{photon, lab=$W$, tens=2}{v1,v2}
\fmf{top, fore=myr, lab=$ $, l.side=right, tens=2}{v1,v3}
\fmf{top, fore=myr}{v3,r1}\fmflabel{$\myr{t}$}{r1}
\fmffreeze
\fmf{gluon, tens=2}{v3,v4}
\fmf{top, fore=myr}{r2,v4,r3}\fmflabel{$\myr{\bar{t}}$}{r2}\fmflabel{$\myr{t}$}{r3}
\end{fmfgraph*}}
\fmfframe(5,5)(5,5){\begin{fmfgraph*}(25,15)
\fmfleft{l1,l2}
\fmfright{r1,r2,r3,r4}
\fmf{fermion,tens=2, fore=myg}{v1,l1}\fmflabel{\myg{$\bar{b}$}}{l1}
\fmf{gluon,tens=2}{l2,v1}\fmflabel{$g$}{l2}
\fmf{fermion, fore=myg, tens=3}{v2,v1}
\fmf{photon}{v2,v8,r4}\fmflabel{$W^+$}{r4}
\fmf{top, fore=myr, lab=$ $, l.side=right, tens=2}{v3,v2}
\fmf{top, fore=myr}{r1,v3}\fmflabel{\myr{$\bar{t}$}}{r1}
\fmffreeze
\fmf{gluon, tens=2, lab=$ $, lab.side=left, lab.d=1mm}{v3,v4}
\fmf{top, fore=myr}{r2,v4,r3}\fmflabel{\myr{$\bar{t}$}}{r2}\fmflabel{\myr{$t$}}{r3}
\end{fmfgraph*}}

\caption{At lowest order, three-top-quark production at the LHC can be thought of as single production ---where three partonic channels can be distinguished--- with the emission of an additional top-antitop quark pair.
}
\label{fig:three-channels}
\end{figure}
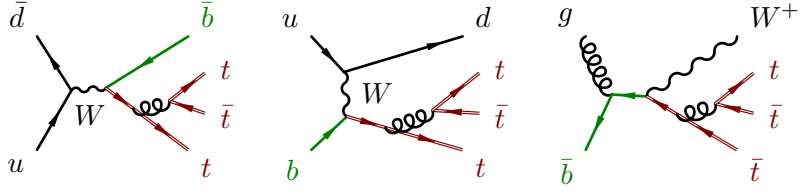

\subsection{Leading-order processes}

At lowest order, three-top-quark production can be conceived as single-top-quark production with the emission of an extra top--anti-top pair (see~\cref{fig:three-channels}).
The classification of single-top production in three partonic channels ($s$-channel, $t$-channel, and $tW$ associated production) can therefore be applied to three-top production too.
The 5FS formally treats together the production of three tops in association with a $b$ or light quark ($s$- and $t$-channel-like), therefore only distinguishing the production of three tops in association with a $W$ boson and with a jet, possibly $b$-flavoured and labelled $j$:
\begin{equation}
\begin{aligned}
&pp \;\longrightarrow\; t\bar{t}tj &\:\text{ and }\:&& \bar{t}t\bar{t}j
\;, \quad\\
&pp \;\longrightarrow\; t\bar{t}tW^{-} &\:\text{ and }\:&& \bar{t}t\bar{t}W^{+}
\;.
\end{aligned}
\label{eq:process}
\end{equation}
At LO, the two charge-conjugate $tttW$ processes are initiated by $gb$ and $g\bar{b}$ partons, therefore having identical rates and kinematics.
In $tttj$ production on the contrary, the two conjugate processes proceed through partonic initial states which have different PDFs.

Unlike in single top-quark production where the $t$-channel contribution is the largest, three-top-quark production is dominated by the $W$-associated production mode, with a LO cross section of the femtobarn (fb) order at the 13~TeV LHC, a factor of 2-3 times (depending on the scale choice) larger than that of the $s$- and $t$-channel modes combined~\cite{Barger:2010uw, Boos:2021yat}.
For comparison, the four-top-quark cross-section is about an order of magnitude larger, in the $\mathcal{O}(10~\text{fb})$ ballpark~\cite{Bevilacqua:2012em, Maltoni:2015ena,  Frederix:2017wme}.
We will therefore focus on the $tttW$ channel and on $\bar{t}t\bar{t}W^{+}$ production in particular.

\subsection{Overlap removal at NLO in QCD}
\label{sec:overlap-removal}
Extra radiation blurs the distinction between three- and four-top production at NLO in QCD.
The same final states can be produced through both resonant and non-resonant diagrams, inducing a so-called channel overlap.
An extra $b$ jet radiated in $tttW$ at NLO can produce the same final state as $tttt$ production at LO with a $t\to Wb$ top decay.\footnote{Similarly, an additional jet emitted in $tttj$ at NLO can lead to the same $tttjj$ final state as $tttW$ production with a hadronic $W\to jj$ decay.}

Strictly speaking, it implies that these different processes cannot be distinguished.
Still, it is convenient to artificially split their joint prediction in two components, each computed at NLO, to achieve the corresponding accuracy in the widest possible range of kinematic configurations.
In the on-shell limit, the resonant diagrams constitute a well-defined first component.
The definition of the second component of the joint prediction is more involved.
Several approaches have been considered in the literature.
Let us discuss them now.

\begin{figure}[tb]\centering
\fmfframe(5,5)(5,5){\begin{fmfgraph*}(25,15)
\fmfleft{l1,l2}
\fmfright{r1,r2,r3,r4,r5}
\fmf{gluon,tens=2}{l2,v2}\fmflabel{$g$}{l2}
\fmf{top, fore=myr, lab=\myr{$t$}, l.side=left, tens=2}{v2,v1}
\fmf{fermion,fore=myg}{v1,r5}\fmflabel{\myg{$b$}}{r5}
\fmf{photon}{v1,r4}\fmflabel{$W^+$}{r4}
\fmf{top, fore=myr, lab=$ $, l.side=right, tens=2}{v5,v3,v2}
\fmf{gluon,tens=3}{l1,v5}\fmflabel{$g$}{l1}
\fmf{top, fore=myr}{r1,v5}\fmflabel{\myr{$\bar{t}$}}{r1}
\fmffreeze
\fmf{gluon, tens=2}{v3,v4}
\fmf{top, fore=myr}{r3,v4,r2}\fmflabel{\myr{$\bar{t}$}}{r3}\fmflabel{\myr{$t$}}{r2}
\end{fmfgraph*}}
\hspace{1cm}%
\fmfframe(5,5)(5,5){\begin{fmfgraph*}(25,15)
\fmfleft{l1,l2}
\fmfright{r1,r2,r3,r4,r5}
\fmf{fermion,fore=myg}{v1,r5}\fmflabel{\myg{$b$}}{r5}
\fmf{fermion, fore=myg, tens=1}{v2,v1}
\fmf{gluon,tens=2}{l2,v1}\fmflabel{$g$}{l2}
\fmf{photon}{v2,v8,r4}\fmflabel{$W^+$}{r4}
\fmf{top, fore=myr, lab=$ $, l.side=right, tens=2}{v5,v3,v2}
\fmf{gluon,tens=3}{l1,v5}\fmflabel{$g$}{l1}
\fmf{top, fore=myr}{r1,v5}\fmflabel{\myr{$\bar{t}$}}{r1}
\fmffreeze
\fmf{gluon, tens=2}{v3,v4}
\fmf{top, fore=myr}{r3,v4,r2}\fmflabel{\myr{$\bar{t}$}}{r3}\fmflabel{\myr{$t$}}{r2}
\end{fmfgraph*}}
\caption{Examples of resonant (left) and non-resonant (right) diagrams contributing to the $tttWb$ final state.
In resonant diagrams, the $Wb$ pair originates from an $s$-channel top-quark propagator which can possibly be on-shell.
In the on-shell limit, these factorise into $pp\to tttt$ production followed by $t\to Wb$ decay.
In non-resonant diagrams, the $Wb$ pair does not originate from a top quark.
In the five-flavour scheme, these correspond to $pp\to tttW$ production with an extra $b$ jet radiation.}
\label{fig:resonant-nonresonant-diagrams}
\end{figure}
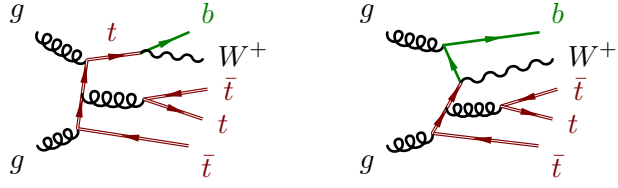

\subsubsection{Diagram removal and subtraction}
The diagram removal (DR) prescriptions distinguish two families of diagrams among those featuring an extra radiated parton (see \cref{fig:resonant-nonresonant-diagrams}).
In resonant diagrams, the extra parton arises from an $s$-channel propagator which can possibly be on-shell.
In non-resonant diagrams, the extra parton never arises from an on-shell resonance.
\begin{itemize}
\item The DR$_1$ prescription~\cite{Frixione:2008yi} excludes resonant diagrams altogether, including only the square of non-resonant ones.
\item Its DR$_2$ variant~\cite{Hollik:2012rc} only excludes the square of resonant diagrams.
In addition to the square of non-resonant diagrams, it therefore also includes their interference with resonant ones.
\end{itemize}
By separating diagrams leading to the same final state, the DR prescriptions in general violate gauge invariance.
The strict on-shell kinematic configuration is the only one in which the resonant diagrams are well-defined in isolation.
The further away one departs from the on-shell region, the more drastic the consequences can be.

Let us illustrate the potential issues in a toy $gt\to bW$ scattering example at high energies.
The two contributing tree-level diagrams, depicted in~\cref{eq:gt2bw}, respectively belong to the resonant and non-resonant categories defined above.\footnote{In the high-energy limit, the fact that the $s$-channel top-quark of the first diagram can actually not be on-shell is immaterial.}
Neglecting mass corrections and taking the $\varepsilon_W^\mu \to p_W^\mu/m_W$ high-energy limit of the longitudinal $W$ polarisation, one obtains:
\begin{equation}
\begin{aligned}
\raisebox{-9mm}{\fmfframe(3,5)(3,5){\begin{fmfgraph*}(25,10)
\fmfleft{l1,l2}
\fmfright{r1,r2}
\fmf{top, fore=myr}{l1,v1}	\fmflabel{\myr{$t$}}{l1}
\fmf{top,tens=2,fore=myr, lab=\myr{$t$}}{v1,v2}
\fmf{fermion, fore=myg}{v2,r2}	\fmflabel{\myg{$b$}}{r2}
\fmf{gluon}{l2,v1}	\fmflabel{$g$}{l2}
\fmf{photon}{v2,r1}	\fmflabel{$W$}{r1}
\end{fmfgraph*}}}
\;&\sim
\frac{\bar{u}_b \slashed{\varepsilon}_W (\slashed{p}_b+\slashed{p}_W) \slashed{\varepsilon}_g u_t}{(p_b+p_W)^2}
\sim +\frac{\bar{u}_b \slashed{\varepsilon}_g u_t}{m_W}
\:,
\\
\raisebox{-9mm}{\fmfframe(3,5)(3,5){\begin{fmfgraph*}(25,10)
\fmfleft{l1,l2}
\fmfright{r1,r2}
\fmf{top, fore=myr}{l1,v1}	\fmflabel{\myr{$t$}}{l1}
\fmf{fermion,tens=0,lab=\myg{$b$}, fore=myg}{v1,v2}
\fmf{fermion, fore=myg}{v2,r2}	\fmflabel{\myg{$b$}}{r2}
\fmf{gluon}{l2,v2}	\fmflabel{$g$}{l2}
\fmf{photon}{v1,r1}	\fmflabel{$W$}{r1}
\end{fmfgraph*}}}
\;&\sim
\frac{\bar{u}_b \slashed{\varepsilon}_g (\slashed{p}_t-\slashed{p}_W) \slashed{\varepsilon}_W u_t}{(p_t-p_W)^2}
\sim -\frac{\bar{u}_b \slashed{\varepsilon}_g u_t}{m_W}
\:.
\end{aligned}
\label{eq:gt2bw}
\end{equation}

In a general gauge for the gluon, both diagrams separately grow linearly with the energy, and this unitarity-violating behaviour is cancelled in their sum.
At the squared amplitude level, it is their interference that grows twice as negative as the individual diagrams squared to restore a well-behaved high-energy behaviour.
In this toy example, retaining only the second diagram squared corresponds to the DR$_1$ prescription and adding to that the interference between both diagrams corresponds to the DR$_2$ prescription.
Both procedures lead to unitarity-violating energy tails, tending to equal and opposite values, positive for DR$_1$ and negative for DR$_2$.
Again, gauge invariance and unitarity are only preserved when all contributions are included.

This is a specific instance of the textbook unitarity violation caused by longitudinal $W$ bosons at high energies, when some contributing diagrams are missed.
Note that a clever choice of gauge may avoid the energy growth of subsets of diagrams.
In the toy example above, choosing an axial gauge for the gluon with a reference momentum aligned with one of the quark momenta can suppress the leading high-energy term by a factor of the fermion mass and therefore tames its growth.
Each diagram, taken in isolation, is still gauge dependent.

To avoid such gauge-invariance and unitarity violation issues, diagram subtraction (DS) schemes can be employed (see e.g.\ \cite{Demartin:2016axk, Frixione:2019fxg} for examples and details).
These procedures locally subtract, from the squared sum of the resonant and non-resonant diagrams, the resonant diagrams squared projected onto a strict on-shell kinematic configuration and multiplied by a Breit-Wigner-like function of the resonance invariant mass.
The specific implementation of such prescriptions however involves a significant amount of arbitrariness, for example in the reshuffling of momenta needed to achieve the strict on-shell condition which ensures gauge invariance.

\subsubsection{Window removal}

Aiming at preserving the simplicity of diagram removal schemes, while avoiding problematic gauge invariance and unitarity violations, we introduce a new variant of diagram removal, dubbed {window removal} and abbreviated DR$_W$.
In the first component of the joint computation, resonant diagrams squared are considered in a finite invariant-mass window, as commonly done to produce realistic events with resonances decayed.
They are well-defined in the strict on-shell limit and a reasonable behaviour is therefore expected for a sufficiently narrow window.
This component of the joint prediction, not discussed further here, is complemented by a second one, which includes all remaining contributions:
\begin{itemize}
\item In the on-shell window, the DR$_W$ prediction includes non-resonant diagrams squared and their interference with resonant ones.
It is therefore identical to the DR$_2$ prescription in that region.
\item Outside of the on-shell window, the DR$_W$ prediction includes resonant and non-resonant diagrams without distinction, which ensures gauge invariance in energy tails where unitarity violation can otherwise occur.
\end{itemize}
The combination of the resonant prediction inside the window with the missing components provided by such a DR$_W$ computation yields a consistent prediction in both on-shell and off-shell regimes.
The motivation to compute these two components separately is, as for all overlap removal prescriptions, to efficiently achieve the highest perturbative accuracy in the widest possible range of kinematic configurations.
In particular, the resonant prediction (with the resonance decay accounted for in the narrow width approximation) is often available at a higher coupling order compared to the non-resonant one.
If both resonant and DR$_W$ components of the joint prediction were computed with the exact same setup and at the same coupling order, the width of the on-shell window would be immaterial.
In practice, these two components of the joint prediction may be computed by different groups, at different orders, in different gauges, maybe with slightly different input parameters.
It is therefore preferable to adopt a window width that is not too large, to reduce the possible impact of residual gauge dependencies, and not too narrow, to model the overwhelming part of the resonant component at the highest possible accuracy and limit the fraction of its lower-order description in the DR$_W$ component of the joint computation.

To avoid double counting, it is crucial to restrict the resonant component of the computation to the invariant-mass window used in the DR$_W$ component.
As the resonant component often has a much larger rate, a small fraction of it can still have a large impact relative to the DR$_W$ one.
This invariant-mass window cut can be implemented on existing event samples by accessing the truth-level invariant mass of resonance decay products.
Given the importance of the on-shell window specification, we will often label the removal procedure as DR$_W^x$, by the employed half-width $x$ around the resonance mass measured in GeV (thus corresponding to a $[m_\text{res}-x\:\text{GeV},m_\text{res}+x\:\text{GeV}]$ window).

\begin{figure}[tb]\centering
\includegraphics[width=.5\textwidth]{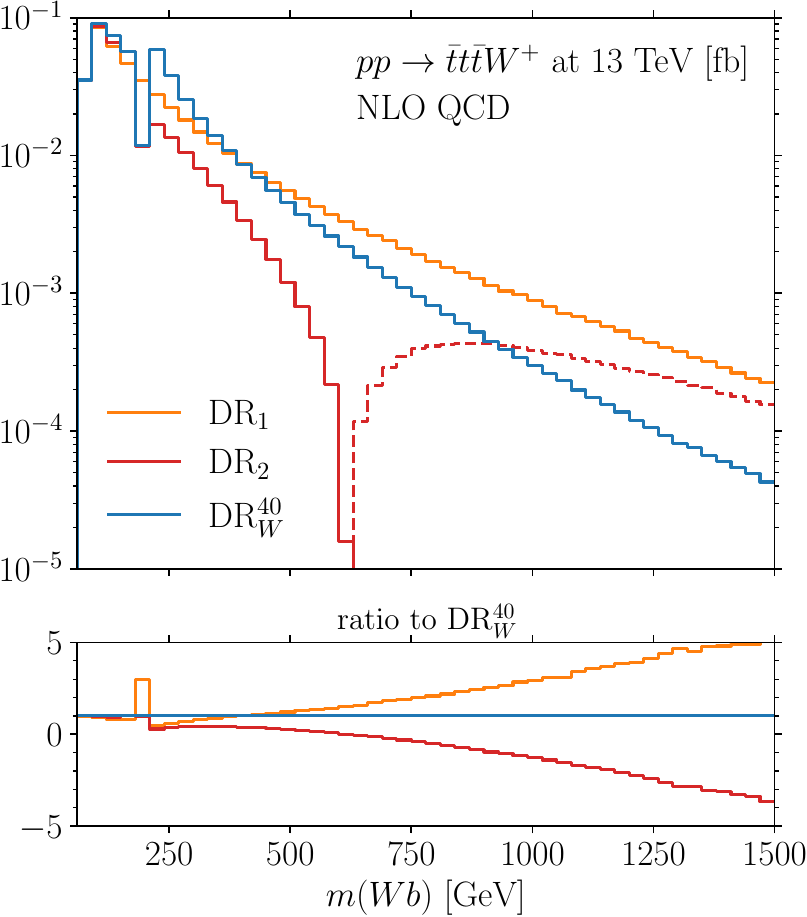}%
\includegraphics[width=.5\textwidth]{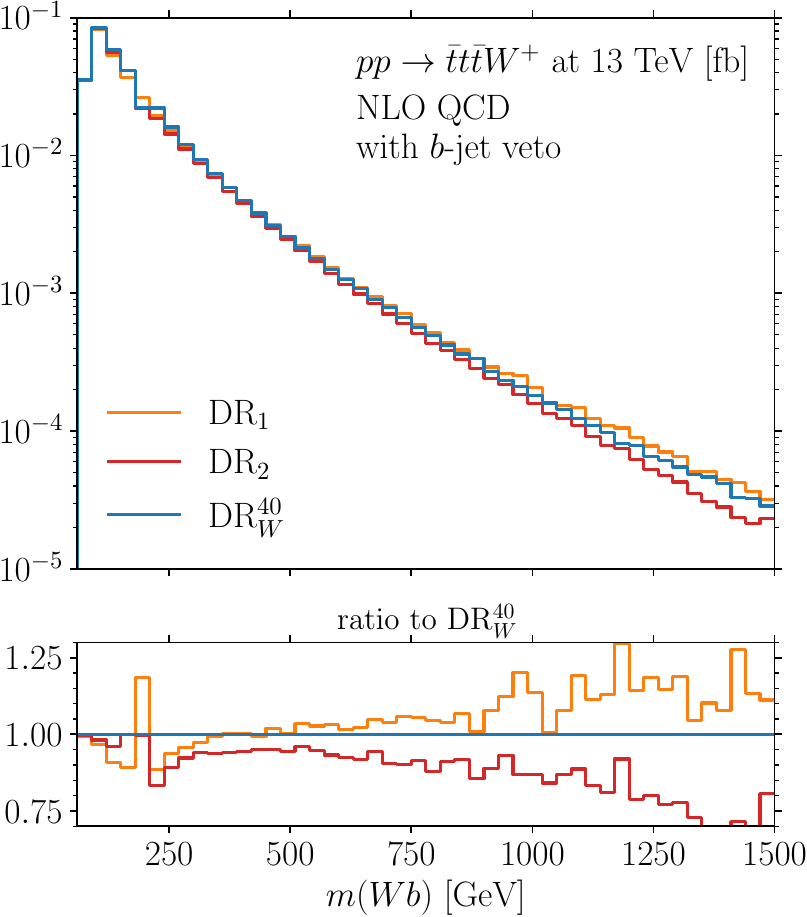}
\caption{Distribution of the invariant mass of the $W$ boson and extra $b$ quark radiated at NLO QCD in $tttW$ production in the 5FS, for different overlap removal prescriptions, without (left panel) and with (right panel) an idealised $b$-jet veto.
Dashed lines indicate negative rates.
}
\label{fig:mwb}
\end{figure}

\subsubsection{The \texorpdfstring{$tttW$}{tttW} case}
Overlap removal procedures are automated in the \texttt{MadSTR} plugin~\cite{Frixione:2019fxg,madstr} of \texttt{MadGraph5\_ aMC@NLO}~\cite{Alwall:2014hca,Frederix:2018nkq}.
We have extended \texttt{MadSTR} so that it can tackle the case of complete-NLO corrections and apply the new DR$_W^x$ family of prescriptions.\footnote{It can be selected by setting the \texttt{ISTR} run-card parameter to $12$ and the \texttt{BWCUTOFF} parameter to the half-width of the desired on-shell window, measured in units of the relevant resonance width (e.g.\ $\texttt{BWCUTOFF}=x\:\text{GeV}/\:\texttt{WT}$ where \texttt{WT} is the top-quark width parameter, if the resonance is a top quark).}
This implementation allows us to examine the realistic case of $tttW$ production with an extra $b$-jet radiated.
Only NLO QCD corrections are considered in this section.
The problematic behaviour of DR$_1$ and DR$_2$ computations described in the toy example above is clearly observed in the tail of the $m({Wb})$ invariant-mass distribution displayed in the left panel of \cref{fig:mwb} (the right panel will be discussed below).
The DR$_1$ and DR$_2$ prescriptions progressively depart from the well-behaved DR$_W$ one at larger $m({Wb})$ values, approximately quadratically and in opposite directions.
The DR$_2$ distribution even turns negative around $m({Wb})\simeq 600$\:GeV.
In the far $m({Wb})$ tail, the DR$_1$ and DR$_2$ computations reach almost equal and opposite values, as expected from the toy example discussed above.
This clearly demonstrates the pathological character of the DR$_1$ and DR$_2$ prescriptions in our setup (with the default \texttt{MadGraph5\_aMC@NLO} gauge choices).

\begin{figure}[tb]\centering
\includegraphics[width=.49\textwidth]{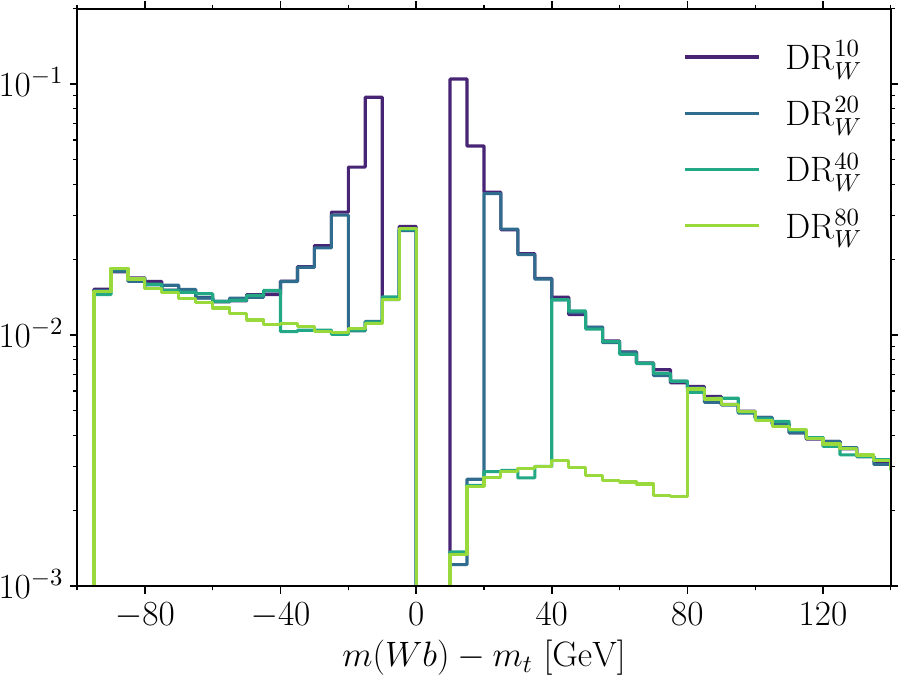}
\caption{Varying the on-shell window size $x$ of the DR$_W^x$ prescription in the $m({Wb})$ distribution of $tttW$ production at NLO QCD in the 5FS.}
\label{fig:window-sizes}
\end{figure}

The impact of varying the on-shell window width of the DR$_W$ prescription for $tttW$ production is displayed in the zoomed-in $m({Wb})$ distribution of \cref{fig:window-sizes}.
The sizeable interference between resonant and non-resonant diagrams gives rise to a peak-dip structure inside the window.
For narrower and narrower windows, the resonant contributions become larger and larger.
A window width of about $\pm40\:$GeV limits these to a level comparable to the interference contributions.
Wider windows increase the phase-space region in which a residual gauge dependence may remain when combining both components of the joint $tttW+tttt$ prediction.
Although other similar values could also be suitable, for definiteness, we will use a $\pm40\:$GeV window size in the following.

\begin{figure}[tb]\centering
\adjustbox{width=\textwidth}{%
\includegraphics{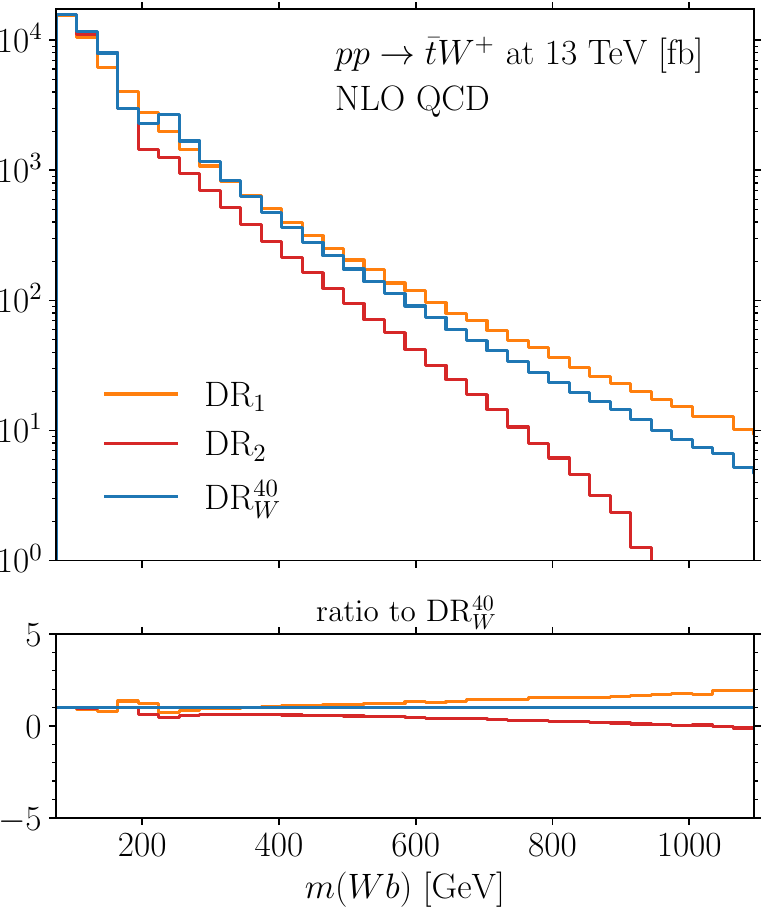}\;\;%
\includegraphics{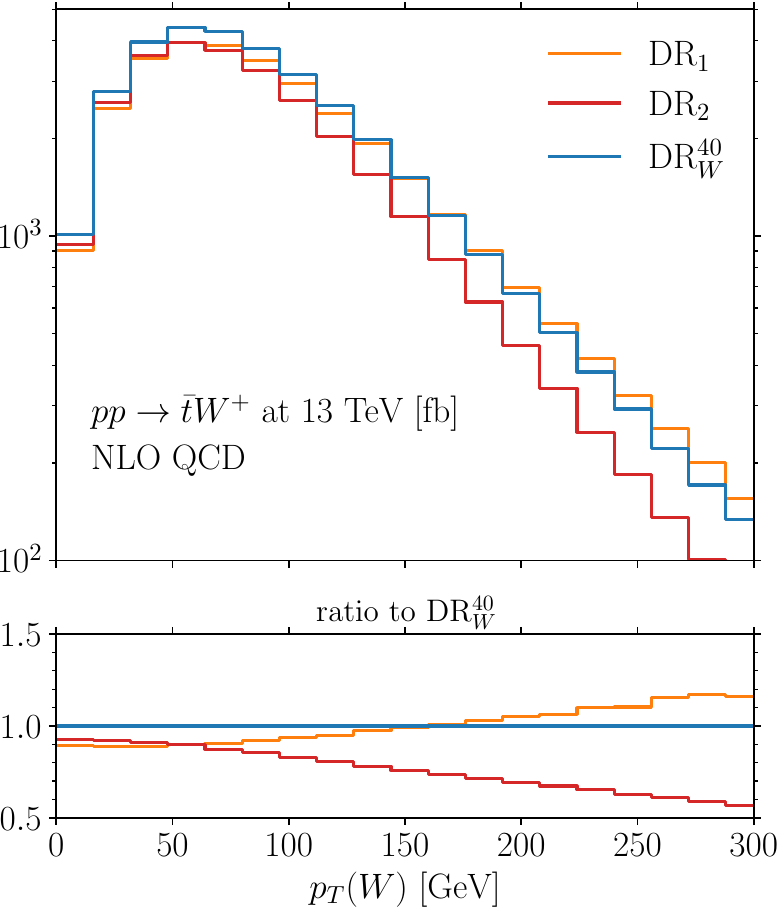}%
}%
\caption{Distribution of the invariant mass of the $W$ boson and extra $b$ quark radiated at NLO in QCD (left) and of the transverse momentum of the $W$ boson (right) in $tW$ production in the 5FS, for different overlap removal prescriptions.
Note the different $y$-axis scales of the ratio panels.
}
\label{fig:mwb-singletop}
\end{figure}

\begin{figure}[tb]\centering
\adjustbox{width=\textwidth}{%
\begin{tabular}{@{}l@{\;}l@{}}%
\includegraphics{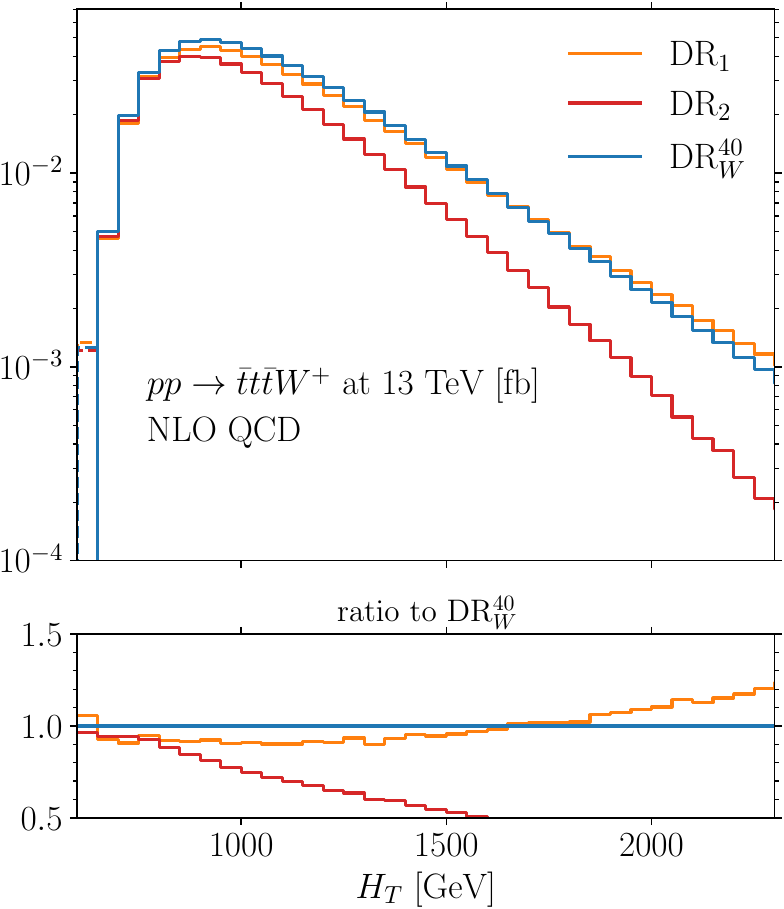}&%
\includegraphics{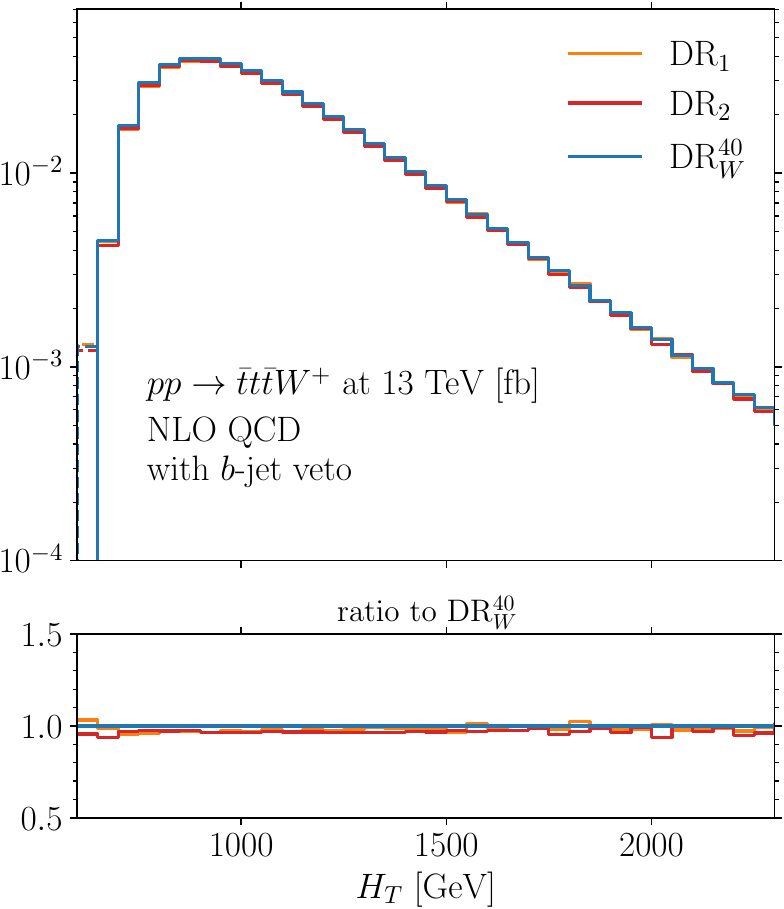}%
\\[5mm]%
\includegraphics{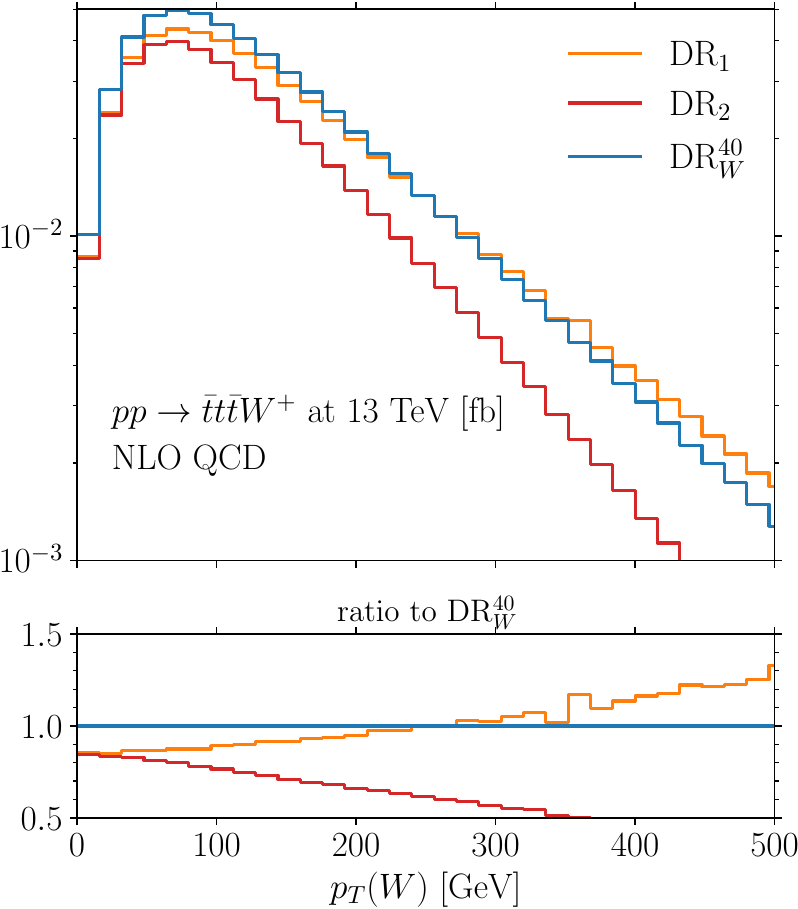}&%
\includegraphics{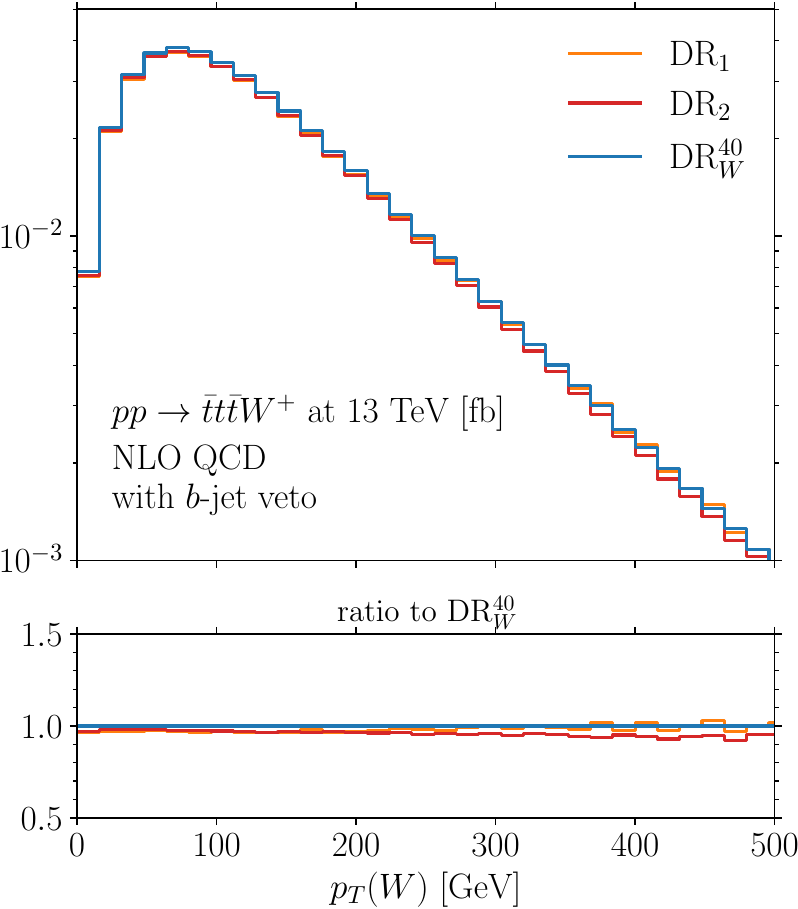}%
\end{tabular}%
}%
\caption{Comparison between the DR$_1$, DR$_2$ and DR$_W$ predictions in illustrative distributions, without (left column) and with (right column) veto on extra $b$-jet radiation.
$H_T$ denotes the scalar sum of transverse energies defined in \cref{eq:ht-def}.
}
\label{fig:dr12w-comparison}%
\end{figure}

Overlap removal procedures have been extensively studied in the literature for $pp\to tW+X$ processes.
In particular, the gauge dependence of DR computations has been studied in detail for $tW$ production in \cite{Frixione:2008yi}.
It was found to have small numerical impact both at the inclusive level and in the moderate regime of energy distributions.
In the left panel of \cref{fig:mwb-singletop}, we show the $m({Wb})$ distribution in $pp\to tW$ production.
In comparison with the left panel of \cref{fig:mwb} obtained with the exact same setup in $tttW$ production, we indeed observe that differences between the DR$_1$, DR$_2$ and DR$_W$ prescriptions are much milder.
The right panel of \cref{fig:mwb-singletop} also shows the transverse momentum of the $W$ boson which can be compared with the distribution of the same variable shown in \cref{fig:dr12w-comparison} for $tttW$ production.
The spread between DR predictions there is comparable to the one observed in $tW+X$ processes~\cite{Frixione:2008yi, Demartin:2016axk, Faham:2021zet}.
The $x$-axis ranges displayed are chosen to span similar logarithmic $y$-axis ranges of cross section per bin in the two processes.

Since the NLO corrections to the partonic $gb\to tttW$ process only include the $gg$- and $b\bar{b}$-initiated $tttt$ production channels, the $q\bar{q}$ ones have no overlap with $tttW$ production.
Therefore, we do not discuss them and do not include their contributions in our figures or tables.
Readily computed at LO, they should however be added to the DR$_W$ prediction, outside of the on-shell window, in comparisons with data.
For reference, outside of the $\pm40\:$GeV window, they amount to about $0.020\:$fb, i.e.\ about $2.8\%$ of the DR$_W$ rate obtained in our setup (see \cref{sec:complete-nlo}).

\paragraph{Idealised \texorpdfstring{$b$}{b} veto}
The DR$_W$ prescription introduced above only makes sense when combined together with a $tttt$ computation inside the on-shell window, to produce a joint $tttW+tttt$ prediction.
Since the $tttW$ and $tttt$ processes are not physically distinct, treating them together is actually sensible.
Their joint prediction can be compared to the data in any desired phase-space region.
Still, to perform more granular tests of the SM, one could wish to define signal regions which are more $tttt$- and $tttW$-like.
To help the construction of such signal regions, two distinct but consistent predictions are needed.

As mentioned several times already, $tttt$ production is well-defined in the strict on-shell limit for the four top quarks.
The associated event samples can readily be generated.
For $tttW$ production, as stated earlier too, a consistent individual definition is possible at LO in the 5FS.
Since small kinematic differences may be critical to gain discrimination power over $tttt$, a more robust NLO prediction however seems desirable.
At that order, one can get inspiration from the definition of a $tW$-like process separate from $t\bar{t}$ production, relying on the veto of a hard and central $b$-jet radiation.
This $b$-jet veto efficiently suppresses the contribution of resonant $t\bar{t}$ diagrams squared and their interference with non-resonant $tW$ ones, since top decays tend to generate harder and more central $b$ jets.
In this $b$-vetoed phase space, the consistent joint $tW+t\bar{t}$ prediction is well approximated by just non-resonant $tW$ diagrams squared.
This therefore provides a phase-space restriction in which the separate $tW$ prediction acquires an effective practical meaning.

In the case of $tttW$ production, being able to distinguish the extra $b$-jet radiation from the three $b$ jets originating from top-quark decays seems very challenging in practice.
Nevertheless, this is possible theoretically in the NLO generation of $tttW$ with three on-shell top quarks.
If the $b$ veto efficiently suppresses the interference with $tttt$ diagrams, event samples simulated in this way could provide a well-defined practical target for the definition of a $tttW$-like signal region, without an explicit $b$-jet veto, in which data could be compared to the consistent $tttW+tttt$ joint prediction.

To test this idea, we veto the radiation of a hard and central $b$ quark with $p_T(b)>30$\:GeV and $|\eta(b)|<2.5$ in $tttW$ production at NLO in QCD.
As can be seen in the right panel of \cref{fig:mwb}, such a veto does suppress very much both the square of resonant $tttt$ diagrams and their interference with non-resonant $tttW$ ones.
The tails of the $m({Wb})$ distribution obtained with the different DR prescriptions indeed agree at the $\mathcal{O}(10\%)$ level until $m({Wb})\simeq 900\:$GeV.
This confirms that this ideal $b$-radiation veto achieves its goal.
For definiteness, in the rest of this paper, we also adopt the DR$_W^{40}$ prescription for predictions with a $b$ veto (see the motivation above for choosing a $\pm40\:$GeV window width).

\paragraph{Comparison between prescriptions}

Let us also compare the different DR prescriptions on distributions other than $m({Wb})$.
The left panels of \cref{fig:dr12w-comparison} show an energy-growing separation between DR$_1$ and DR$_2$ predictions in two energy distributions.
This effect is especially marked in the distribution of the transverse momentum of the $W$ boson, particularly correlated with $m({Wb})$.
In the opposite regime, at lower energies, DR$_1$ and DR$_2$ tend to converge to a rate about $10\%$ lower than the DR$_W^{40}$ one.
The DR$_W^{40}$ prediction indeed includes $tttt$ diagrams squared outside of the on-shell window, which are altogether absent in both DR$_1$ and DR$_2$ predictions.
In a proper joint $tttW+tttt$ prediction, this offset would be compensated by the resonant $tttt$ component that is not restricted to an on-shell window in the DR$_1$ and DR$_2$ cases.

The right panels of \cref{fig:dr12w-comparison} show that enforcing a $b$-jet veto makes the DR$_1$, DR$_2$ and DR$_W$ predictions converge, approximately to the $3\%$ level.
Again, the DR$_W$ prediction tends to be slightly higher than the DR$_1$ and DR$_2$ ones, due to the contributions of $tttt$ diagrams just outside of the on-shell window.

Given the pathologies they exhibit, we will not consider the DR$_1$ and DR$_2$ prescriptions further, using solely the DR$_W$ one with or without $b$ radiation veto.

\section{Complete NLO prediction for \texorpdfstring{$tttW$}{tttW}}
\label{sec:complete-nlo}

We are now equipped with the needed prescriptions to compute $tttW$ production beyond the LO, either as an inseparable component of a joint $tttW+tttt$ prediction using the DR$_W$ procedure, or in a restricted $b$-vetoed phase-space where it is separately well-defined thanks to a suppression of its interference with $tttt$ production.

\subsection{Setup and notation}
\label{sec:setup}
We perform fixed-order calculations for $pp\to \bar{t}t\bar{t}W^+$ production at a centre-of-mass energy of \(\sqrt{s}=13\ \mathrm{TeV}\), with \texttt{MadGraph5\_aMC@NLO}~\cite{Alwall:2014hca, Frederix:2018nkq}. 
We employ the $G_\mu$ scheme for the electroweak input parameters,
with $G_\mu = 1.16639\times10^{-5}\ \mathrm{GeV}^{-2}$, $m_Z = 91.188\ \mathrm{GeV}$, $m_W = 80.419\ \mathrm{GeV}$, and the corresponding renormalisation conditions.
The top-quark mass is set to \(m_{t}=174.3\ \mathrm{GeV}\), whilst decay widths are set to zero.\footnote{Technically, a non-zero decay width for the top quark needs to be kept only in the resonant diagrams to regulate the on-shell integrable divergence of the matrix element.
The results after removal of the on-shell overlap do not depend on its specific value. 
We set it to $1.468\:$GeV.}
We employ the NNPDF3.1 NLO PDF set~\cite{Ball:2017nwa} in the 5FS with \(\alpha_{s}(m_{Z})=0.118\), including photon contributions \emph{\`a la} \texttt{LuxQED}~\cite{Manohar:2016nzj, Manohar:2017eqh}.
It is accessed via the \texttt{LHAPDF} interface~\cite{Buckley:2014ana} (where it is labelled \texttt{NNPDF31\_nlo\_as\_0118\_luxqed} or 324900).
Unless otherwise mentioned, we do not impose any cuts.
When it is applied, the $b$ veto excludes extra hard and central $b$-quark radiation with $p_T(b)>30$\:GeV and $|\eta(b)|<2.5$.

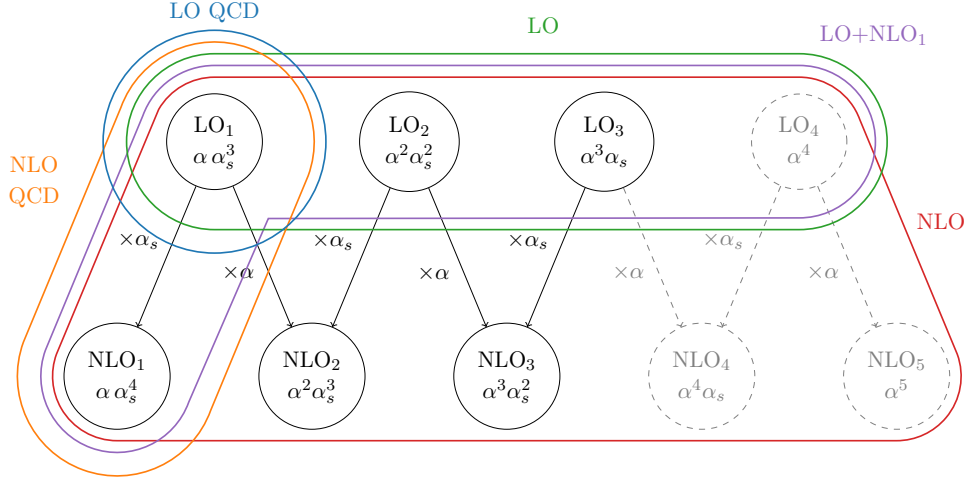
\begin{figure}[tb]\centering
\definecolor{c1}{rgb}{0.12156862745098039, 0.4666666666666667, 0.7058823529411765}
\definecolor{c2}{rgb}{1.0, 0.4980392156862745, 0.054901960784313725}
\definecolor{c3}{rgb}{0.17254901960784313, 0.6274509803921569, 0.17254901960784313}
\definecolor{c4}{rgb}{0.8392156862745098, 0.15294117647058825, 0.1568627450980392}
\definecolor{c5}{rgb}{0.5803921568627451, 0.403921568627451, 0.7411764705882353}
\definecolor{c6}{rgb}{0.5490196078431373, 0.33725490196078434, 0.29411764705882354}
\definecolor{c7}{rgb}{0.8901960784313725, 0.4666666666666667, 0.7607843137254902}
\definecolor{c8}{rgb}{0.4980392156862745, 0.4980392156862745, 0.4980392156862745}
\definecolor{c9}{rgb}{0.7372549019607844, 0.7411764705882353, 0.13333333333333333}
\definecolor{c10}{rgb}{0.09019607843137255, 0.7450980392156863, 0.8117647058823529}
\adjustbox{max width=.85\textwidth}{%
\begin{tikzpicture}
\node (nlo1) [circle, draw] {\minitab{NLO$_1$\\$\alpha\,\alpha_s^4$}};
\node (nlo2) [right=1.5cm of nlo1, circle, draw]{\minitab{NLO$_2$\\$\alpha^2\alpha_s^3$}};
\node (nlo3) [right=1.5cm of nlo2, circle, draw]{\minitab{NLO$_3$\\$\alpha^3\alpha_s^2$}};
\node[gray,dashed] (nlo4) [right=1.5cm of nlo3, circle, draw]{\minitab{NLO$_4$\\$\alpha^4\alpha_s$}};
\node[gray,dashed] (nlo5) [right=1.5cm of nlo4, circle, draw]{\minitab{NLO$_5$\\$\alpha^5$}};
\node (lo1) at ($(nlo1)!.5!(nlo2)$) [yshift=4cm, circle, draw] {\minitab{LO$_1$\\$\alpha\,\alpha_s^3$}};
\node (lo2) at ($(nlo2)!.5!(nlo3)$) [yshift=4cm, circle, draw] {\minitab{LO$_2$\\$\alpha^2\alpha_s^2$}};
\node (lo3) at ($(nlo3)!.5!(nlo4)$) [yshift=4cm, circle, draw] {\minitab{LO$_3$\\$\alpha^3\alpha_s$}};
\node[gray,dashed] (lo4) at ($(nlo4)!.5!(nlo5)$) [yshift=4cm, circle, draw] {\minitab{LO$_4$\\$\alpha^4$}};
\draw[->] (lo1)--node[above left]{$\times\alpha_s$}(nlo1);
\draw[->] (lo2)--node[above left]{$\times\alpha_s$}(nlo2);
\draw[->] (lo3)--node[above left]{$\times\alpha_s$}(nlo3);
\draw[gray, dashed,->] (lo4)--node[above left]{$\times\alpha_s$}(nlo4);
\draw[->] (lo1)--node[below left]{$\times\alpha$}(nlo2);
\draw[->] (lo2)--node[below left]{$\times\alpha$}(nlo3);
\draw[gray,dashed,->] (lo3)--node[below left]{$\times\alpha$}(nlo4);
\draw[gray,dashed,->] (lo4)--node[below left]{$\times\alpha$}(nlo5);
\draw[c4, thick] ($(lo1)+(150:11mm)$) arc [start angle=150, end angle=90, radius=11mm] -- ($(lo4)+(0mm,11mm)$)  arc [start angle=90, end angle=20, radius=11mm] -- node[above right]{NLO}  ($(nlo5)+(20:11mm)$)  arc [start angle=20, end angle=-90, radius=11mm] -- ($(nlo1)+(0mm,-11mm)$) arc [start angle=-90, end angle=-200, radius=11mm] -- cycle;
\draw[c5, thick] ($(lo1)+(160:13mm)$) arc[start angle=160, end angle=90, radius=13mm] -- ($(lo4)+(90:13mm)$) node[above right=3mm]{LO$+$NLO$_1$}  arc[start angle=90, end angle=-90, radius=13mm] --  ($(lo1)+(-55:16mm)$) -- ($(nlo1)+(-20:13mm)$)  arc[start angle=-20, end angle=-200, radius=13mm] -- cycle;
\draw[c3, thick] ($(lo1)+(0mm,15mm)$) --  node[above right=3mm]{LO} ($(lo4)+(0mm,15mm)$)  arc[start angle=90, end angle=-90, radius=15mm] --  ($(lo1)+(0mm,-15mm)$) arc[start angle=-90, end angle=-270, radius=15mm] -- cycle;
\draw[c2, thick] ($(lo1)+(160:17mm)$) arc [start angle=160, end angle=-20, radius=17mm] -- ($(nlo1)+(-20:17mm)$) arc [start angle=-20, end angle=-200, radius=17mm]  --  node[above left=1mm]{\minitab{NLO\\QCD}} cycle;
\draw[c1, thick] (lo1) circle (19mm) node[above=19mm]{LO QCD}; 
\end{tikzpicture}}
\caption{Leading and next-to-leading coupling orders in $tttW$ production in the 5FS, at the squared amplitude level.
Denominations for the various predictions.
The dashed grey LO$_4$, NLO$_4$ and NLO$_5$ contributions only give percent-level corrections and will therefore be neglected.
}%
\label{fig:orders}%
\end{figure}%

\begin{figure}[tb]\centering%
\fmfframe(5,5)(5,5){\begin{fmfgraph*}(25,15)
\fmfleft{l1,l2}
\fmfright{r1,r2,r3,r4}
\fmf{fermion,tens=2, fore=myg}{v1,l1}\fmflabel{\myg{$\bar{b}$}}{l1}
\fmf{gluon,tens=2}{l2,v1}\fmflabel{$g$}{l2}
\fmf{fermion, fore=myg, tens=3}{v2,v1}
\fmf{photon}{v2,v8,r4}\fmflabel{$W^+$}{r4}
\fmf{top, fore=myr, lab=$ $, l.side=right, tens=2}{v3,v2}
\fmf{top, fore=myr}{r1,v3}\fmflabel{\myr{$\bar{t}$}}{r1}
\fmffreeze
\fmf{gluon, tens=2, lab=$ $, lab.side=left, lab.d=1mm}{v3,v4}
\fmf{top, fore=myr}{r2,v4,r3}\fmflabel{\myr{$\bar{t}$}}{r2}\fmflabel{\myr{$t$}}{r3}
\end{fmfgraph*}}
\fmfframe(5,5)(5,5){\begin{fmfgraph*}(25,15)
\fmfleft{l1,l2}
\fmfright{r1,r2,r3,r4}
\fmf{fermion,tens=2, fore=myg}{v1,l1}\fmflabel{\myg{$\bar{b}$}}{l1}
\fmf{photon,tens=2}{l2,v1}\fmflabel{$\gamma$}{l2}
\fmf{fermion, fore=myg, tens=3}{v2,v1}
\fmf{photon}{v2,v8,r4}\fmflabel{$W^+$}{r4}
\fmf{top, fore=myr, lab=$ $, l.side=right, tens=2}{v3,v2}
\fmf{top, fore=myr}{r1,v3}\fmflabel{\myr{$\bar{t}$}}{r1}
\fmffreeze
\fmf{gluon, tens=2, lab=$ $, lab.side=left, lab.d=1mm}{v3,v4}
\fmf{top, fore=myr}{r2,v4,r3}\fmflabel{\myr{$\bar{t}$}}{r2}\fmflabel{\myr{$t$}}{r3}
\end{fmfgraph*}}
\fmfframe(5,5)(5,5){\begin{fmfgraph*}(25,15)
\fmfleft{l1,l2}
\fmfright{r1,r2,r3,r4}
\fmf{fermion,tens=2, fore=myg}{v1,l1}\fmflabel{\myg{$\bar{b}$}}{l1}
\fmf{gluon,tens=2}{l2,v1}\fmflabel{$g$}{l2}
\fmf{fermion, fore=myg, tens=3}{v2,v1}
\fmf{photon}{v2,v8,r4}\fmflabel{$W^+$}{r4}
\fmf{top, fore=myr, lab=$ $, l.side=right, tens=2}{v3,v2}
\fmf{top, fore=myr}{r1,v3}\fmflabel{\myr{$\bar{t}$}}{r1}
\fmffreeze
\fmf{dashes, tens=2, lab=$H$, lab.side=left, lab.d=.8mm}{v3,v4}
\fmf{top, fore=myr}{r2,v4,r3}\fmflabel{\myr{$\bar{t}$}}{r2}\fmflabel{\myr{$t$}}{r3}
\end{fmfgraph*}}
\fmfframe(5,5)(5,5){\begin{fmfgraph*}(25,15)
\fmfleft{l1,l2}
\fmfright{r1,r2,r3,r4}
\fmf{fermion,tens=2, fore=myg}{v1,l1}\fmflabel{\myg{$\bar{b}$}}{l1}
\fmf{photon,tens=2}{l2,v1}\fmflabel{$\gamma$}{l2}
\fmf{fermion, fore=myg, tens=3}{v2,v1}
\fmf{photon}{v2,v8,r4}\fmflabel{$W^+$}{r4}
\fmf{top, fore=myr, lab=$ $, l.side=right, tens=2}{v3,v2}
\fmf{top, fore=myr}{r1,v3}\fmflabel{\myr{$\bar{t}$}}{r1}
\fmffreeze
\fmf{dashes, tens=2, lab=$H$, lab.side=left, lab.d=.8mm}{v3,v4}
\fmf{top, fore=myr}{r2,v4,r3}\fmflabel{\myr{$\bar{t}$}}{r2}\fmflabel{\myr{$t$}}{r3}
\end{fmfgraph*}}

\caption{Representative Feynman diagrams for $tttW$ production at LO, in the five-flavour scheme.
When squared, they contribute to \(\mathrm{LO}_1\), \(\mathrm{LO}_2\), \(\mathrm{LO}_3\), and \(\mathrm{LO}_4\), respectively, from left to right.}
\label{fig:LO_tttW}
\end{figure}
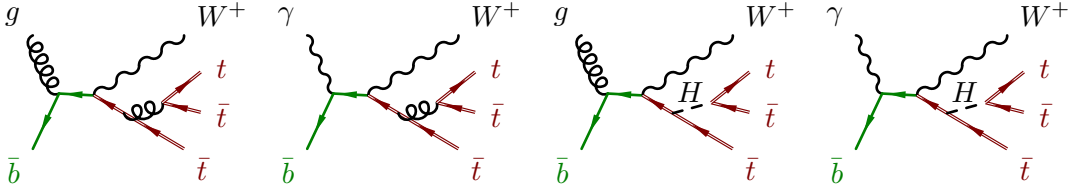

Following the notation of \cite{Frixione:2014qaa, Frixione:2015zaa, Pagani:2016caq, Frederix:2016ost, Czakon:2017wor, Frederix:2017wme, Frederix:2018nkq, Broggio:2019ewu, Frederix:2019ubd, Pagani:2020rsg, Pagani:2020mov, Pagani:2021iwa, ElFaham:2024egs},
the partonic cross section for $tttW$ production can be separated in contributions of different $\alpha^{\,i}\,\alpha_s^{\,j}$ orders (see \cref{fig:orders}).
Four leading orders are present, characterised by $i+j=4$ with $i\ge1$ and $j\ge0$.
They are denoted $\text{LO}_i$ with $i=1,..,4$  and each corresponds to contributions of $\mathcal O(\alpha^i \alpha_s^{4-i} $).
Representative diagrams contributing to these various orders are shown in~\cref{fig:LO_tttW}.
When looking at this figure, the reader should keep in mind that these orders are defined at the level of squared amplitudes.
Whilst it is true that the square of the $i$-th diagram contributes to LO$_i$, interferences also occur.
At NLO, one can either add one power of $\alpha_s$ or one power of $\alpha$ to each LO contribution.
Thus, we end up with five different NLO coupling orders denoted $\text{NLO}_{i}$ with $i=1,..,5$, each corresponding to contributions of $\mathcal O(\alpha^i\alpha_s^{5-i} )$.
All but the extremal NLO$_i$ receive contributions that stem both from QCD corrections to LO$_i$ and from EW corrections to LO$_{i-1}$.
These two classes of contributions cannot be computed separately, lest infrared divergences remain in the cross section.

Coupling powers entering the various contributions suggest that a hierarchy exists among them, with the $(i+1)$-th order being suppressed by a factor of $\alpha/\alpha_s$ with respect to the $i$-th one.
However, this naive expectation can be violated, and is actually strongly violated in the case of $tttt$ production~\cite{Frederix:2017wme}.
The main reason why this happens is that the top-quark Yukawa coupling, despite being of electroweak nature, has an $\mathcal O (1)$ value.
Hence, contributions of tens of percent arise in $tttt$ production from orders up to LO$_3$ and NLO$_4$.
Moreover, these have different signs and large accidental cancellations occur among them.
Since $tttW$ production features diagrams where an internal Higgs boson is exchanged, one can expect a similar pattern.

Contributions from different coupling orders have to be summed to obtain genuine predictions for the process of interest.
We consider five different such predictions in the following (see \cref{fig:orders}):\footnote{When comparing this work with \cite{Frederix:2017wme}, the reader should notice that ${\rm NLO}$ is defined differently.}
\begin{equation}
\begin{aligned}[t]
\LOQCD	&\equiv {\rm LO}_{1}\,, \\
\NLOQCD	&\equiv  {\rm LO}_{1}+{\rm NLO}_{1}
	\,,  
\end{aligned}
\qquad
\begin{aligned}[t]
{\rm LO}&\equiv
	\text{LO}_{1+2+3\textcolor{gray}{+4}}
	\,,\\
\text{LO}+\text{NLO}_1
	&\equiv \text{LO}_{1+2+3\textcolor{gray}{+4}} + \text{NLO}_{1}
	\,,\\
{\rm NLO}&\equiv
	\text{LO}_{1+2+3\textcolor{gray}{+4}} + \text{NLO}_{1+2+3\textcolor{gray}{+4+5}}
	\,,
\label{eq:sec4_notation}
\end{aligned}
\end{equation}
where the shorthand notation $\text{(N)LO}_{i+j+\ldots} = \text{(N)LO}_i + \text{(N)LO}_j + \ldots$ has been introduced.
Note however that the (greyed) contributions from LO$_4$, NLO$_4$ and NLO$_5$ are numerically negligible in $tttW$.
We will therefore neglect them in the following.

\subsection{Scale choices at NLO in QCD}
\begin{figure}[tb]\scriptsize
\begin{tabular}{@{}c@{}}
DR$_W^{40}$\\
\makecell[c]{\quad
	\LOQCD\\\includegraphics[width=.33\textwidth]{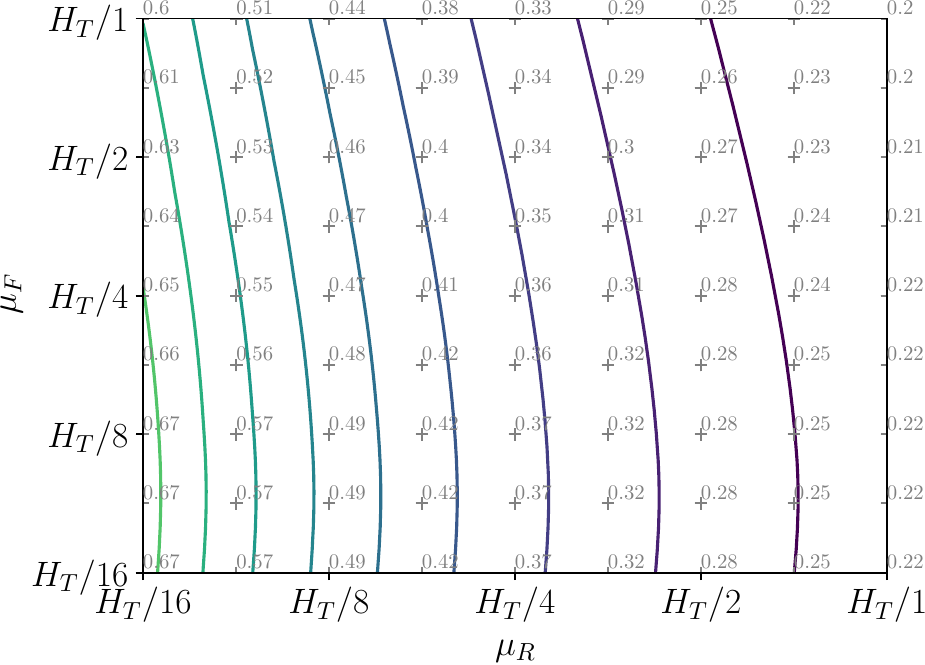}}%
\makecell[c]{\quad
	\NLOQCD%
	\\\includegraphics[width=.33\textwidth]{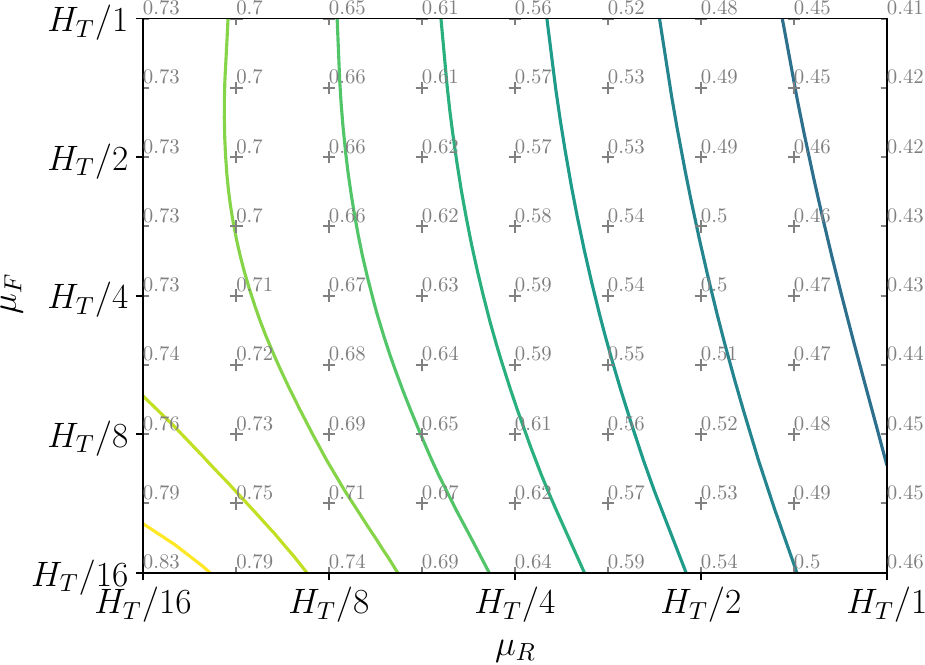}}%
\makecell[c]{\quad
	$K$-factor%
	\\\includegraphics[width=.33\textwidth]{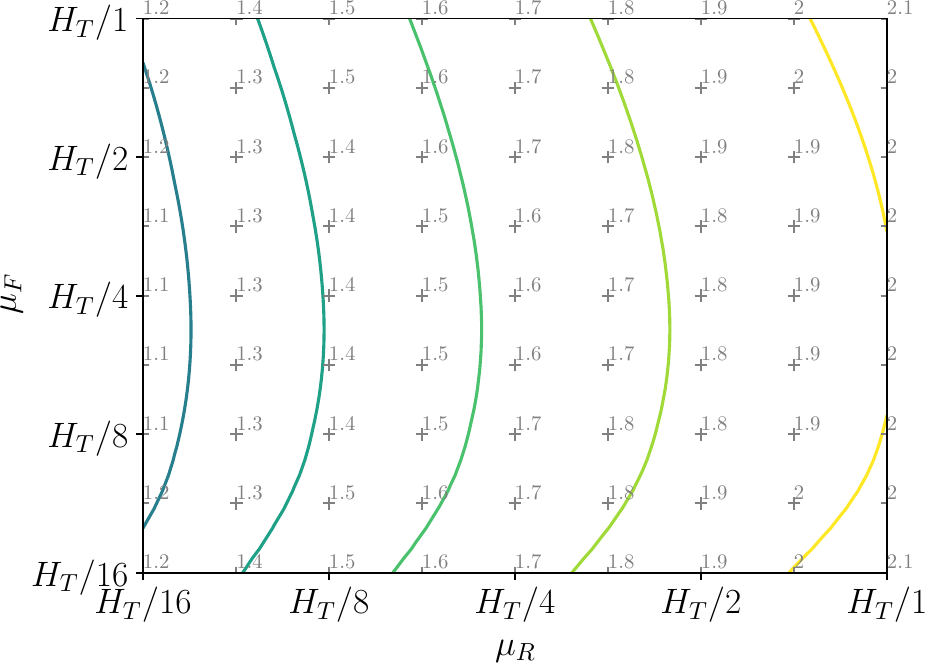}}%
\\[-2mm]
\\
DR$_W^{40}$ + $b$-veto\\
\makecell[c]{\quad
	\LOQCD\\\includegraphics[width=.33\textwidth]{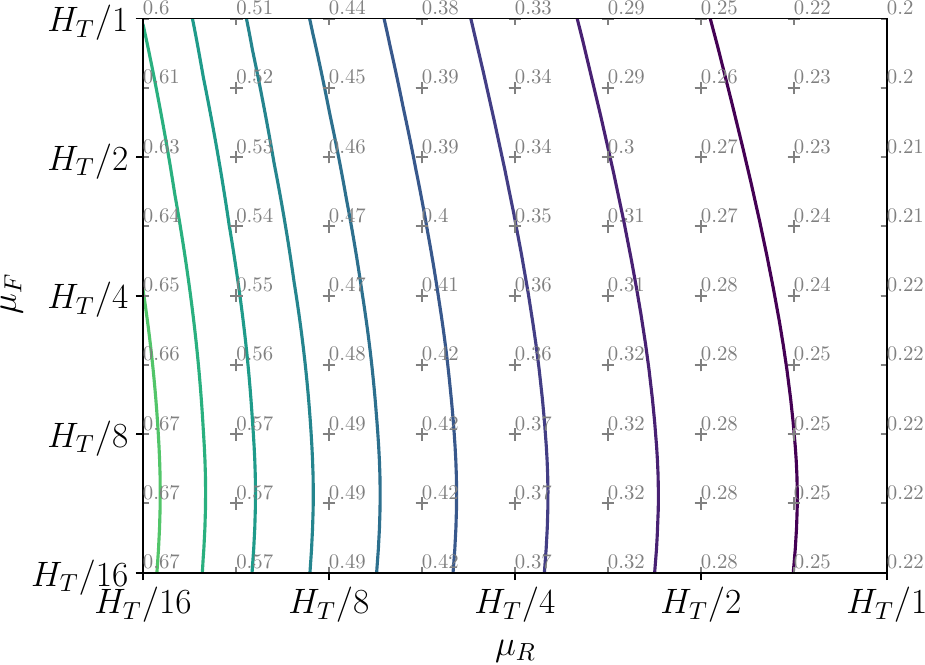}}%
\makecell[c]{\quad
	\NLOQCD%
	\\\includegraphics[width=.33\textwidth]{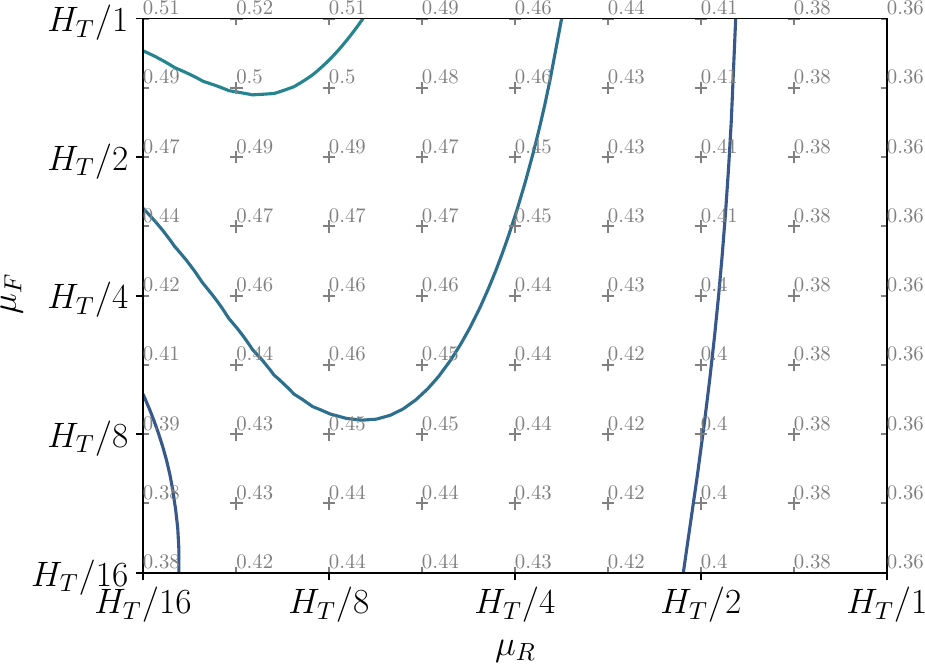}}%
\makecell[c]{\quad
	$K$-factor%
	\\\includegraphics[width=.33\textwidth]{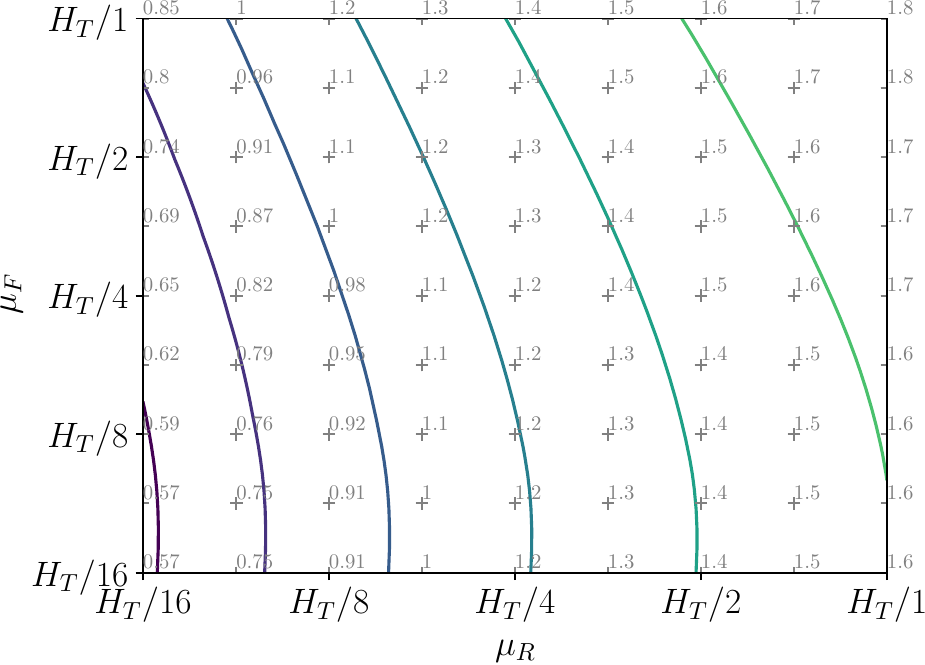}}%
\end{tabular}
\caption{Two-dimensional scale variation for $tttW$ production at the 13~TeV LHC: at lowest order (left column), adding QCD corrections (middle column), and on the ratio between these two ($K$-factor, right column).
NLO QCD results are obtained with the DR$_W^{40}$ prescription, without (top row) and with (bottom row) veto on the radiation of an extra $b$ jet.
Cross sections are given in fb.}
\label{fig:2d-scales}
\end{figure}

Before further explorations, let us first discuss the choices of central renormalisation ($\mu_R$) and factorisation ($\mu_F$) scales for the $tttW$ process at NLO in QCD.
We consider dynamical scales to better model differential distributions and, for simplicity, focus on $2^k$ with $k\in\mathbb{Z}$ multiples of the scalar sum of the transverse energies of all final-state particles, including any extra radiation,
\begin{equation}
H_{T}\equiv \sum_{i\in \text{final}}\sqrt{m_i^2+p_T(i)^2}
\:.
\label{eq:ht-def}
\end{equation}
Illustrative two-dimensional variations are shown in~\cref{fig:2d-scales}, where the main $\mu_R=\mu_F$ diagonal is seen to capture most variations of the \NLOQCD\ rate.
We find no particular motivation for considering a central choice away from this $\mu_R=\mu_F$ diagonal.

\begin{figure}[t]
\hfill
\begin{tabular}{@{}c@{}}
\quad DR$_W^{40}$\\%
\includegraphics[width=.425\textwidth]{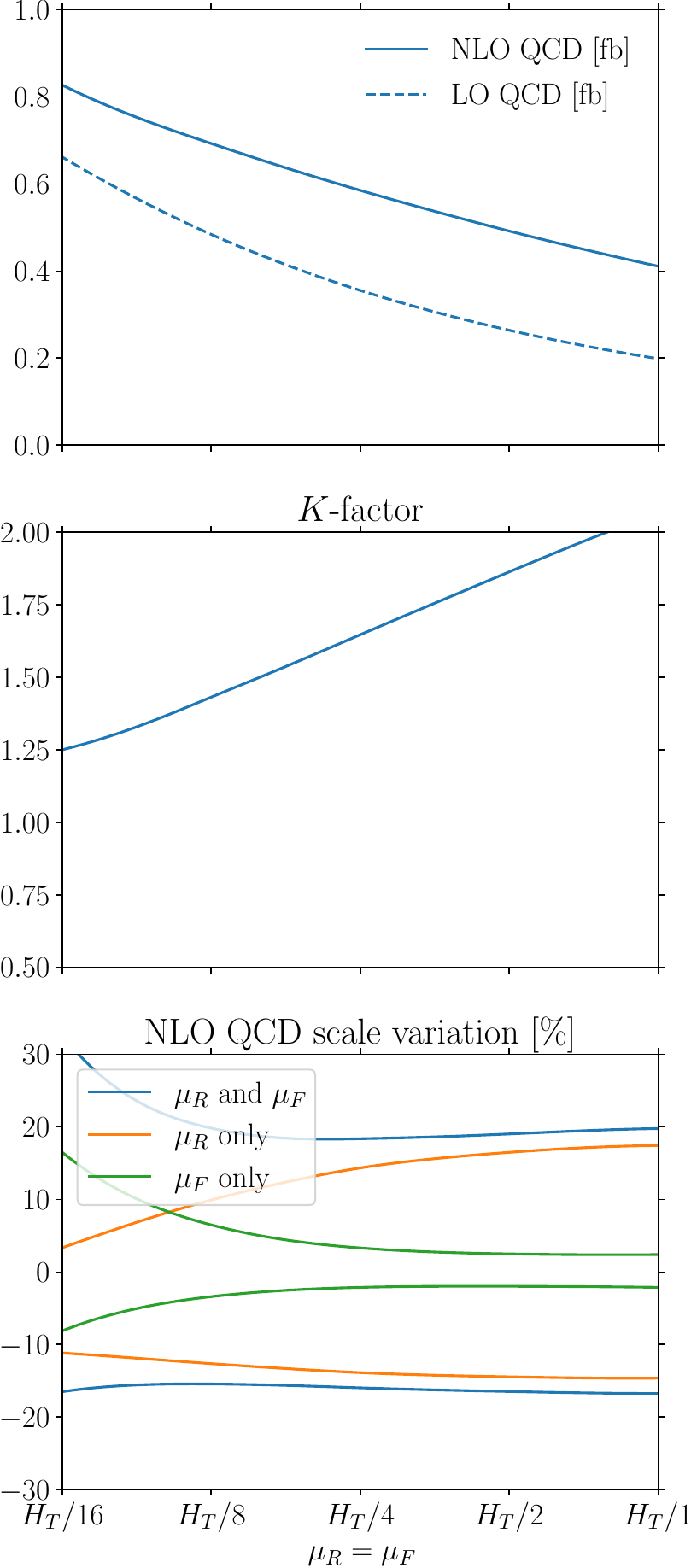}%
\end{tabular}%
\hfill%
\begin{tabular}{@{}c@{}}
\quad DR$_W^{40}$ + $b$-veto\\%
\includegraphics[width=.425\textwidth]{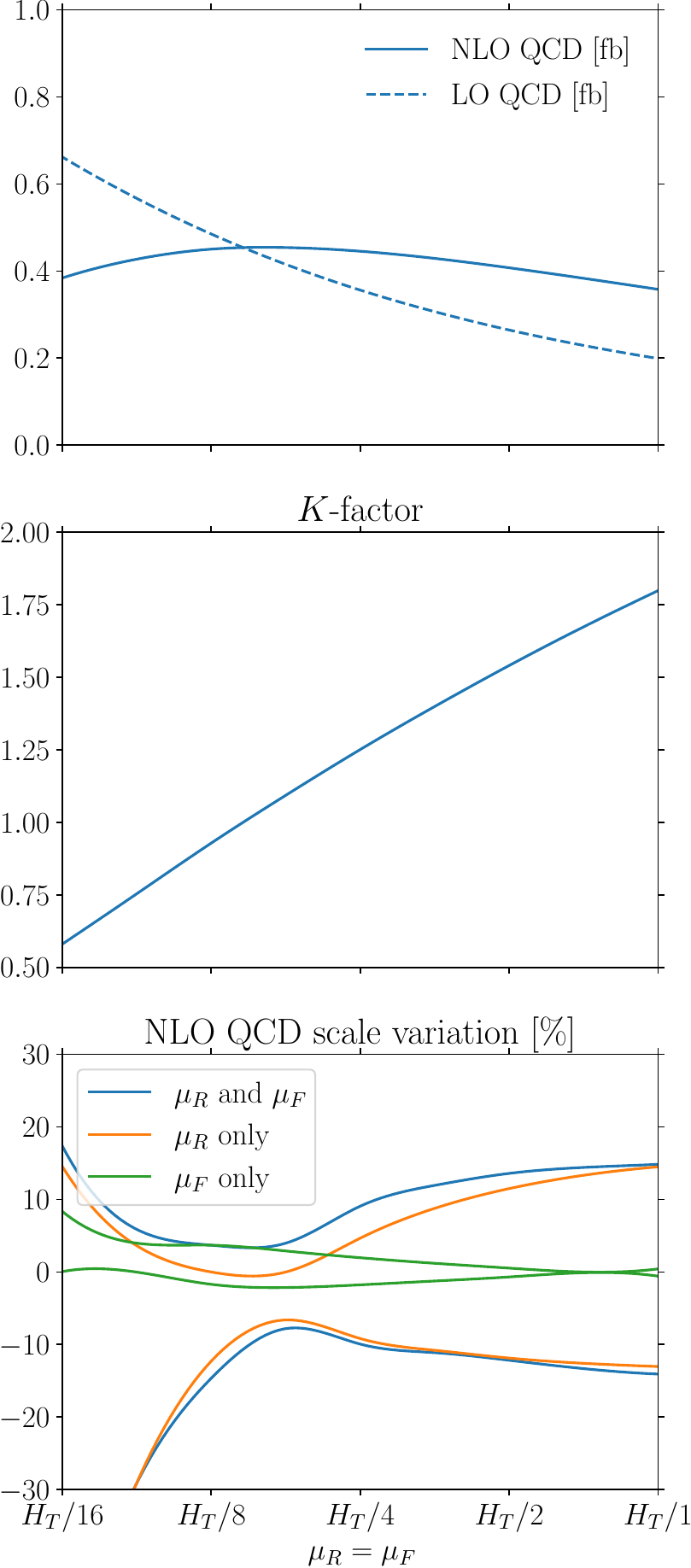}%
\end{tabular}%
\hspace*{\fill}%
\caption{LO QCD and \NLOQCD\ rates (top row), $K$-factors (middle row) and nine-point scale-variation uncertainties (bottom row) as functions of $\mu_R=\mu_F$, for $tttW$ production at the 13~TeV LHC, computed using the DR$_W^{40}$ prescription without (left column) and with (right column) veto on extra $b$-jet radiation.
}
\label{fig:scale-var-diag}
\end{figure}

The perturbative expansion is likely to be best behaved where $K$-factors and factor-of-two scale-variation uncertainties are both moderate.
Those two quantities are shown in \cref{fig:scale-var-diag} as functions of $\mu_R=\mu_F$.
On the one hand, $K$-factors increase for larger scales (first panel).
On the other hand, scale variation uncertainties increase for smaller central scales.
Avoiding either extremes leads to central scale choices around $H_T/4$ (second panel).
In this intermediate regime, individual renormalisation and factorisation scale uncertainties are also of comparable magnitudes (third panel).

Since the $b$-jet veto reduces leading-order $tttt$ contributions as well as their interference with $tttW$, the inclusive rate, NLO QCD $K$-factor, and scale variation uncertainties are somewhat smaller than without veto.

For definiteness, we choose $H_T/4$ as central scale for the $tttW$ production process.

\subsection{Inclusive rates}
\begin{figure}[tb]
\begin{tabular}{@{}c@{\:}c@{}}
\qquad DR$_W^{40}$	& \qquad DR$_W^{40}$ + $b$-veto\\
\includegraphics[width=.5\textwidth]{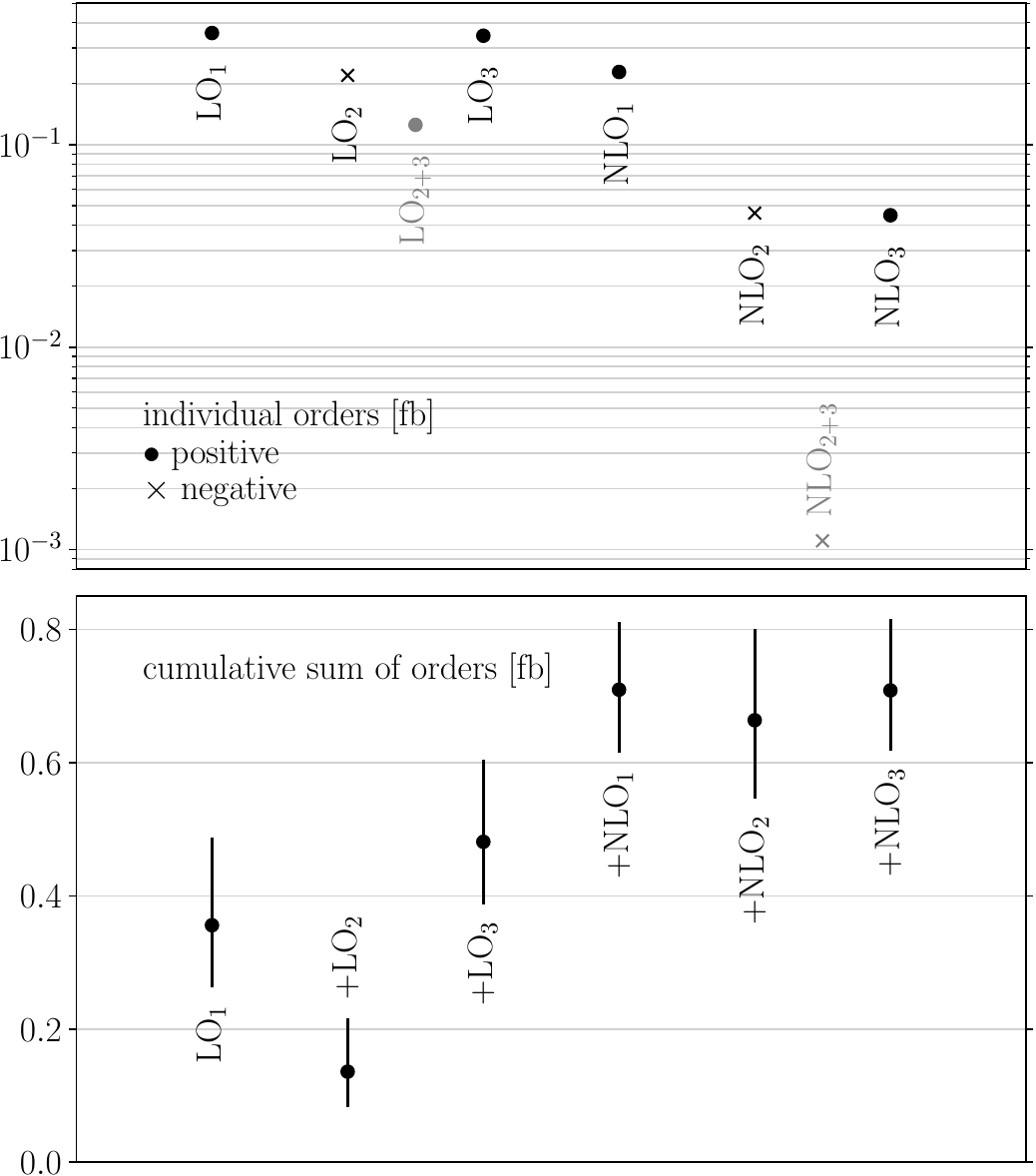}&
\includegraphics[width=.5\textwidth]{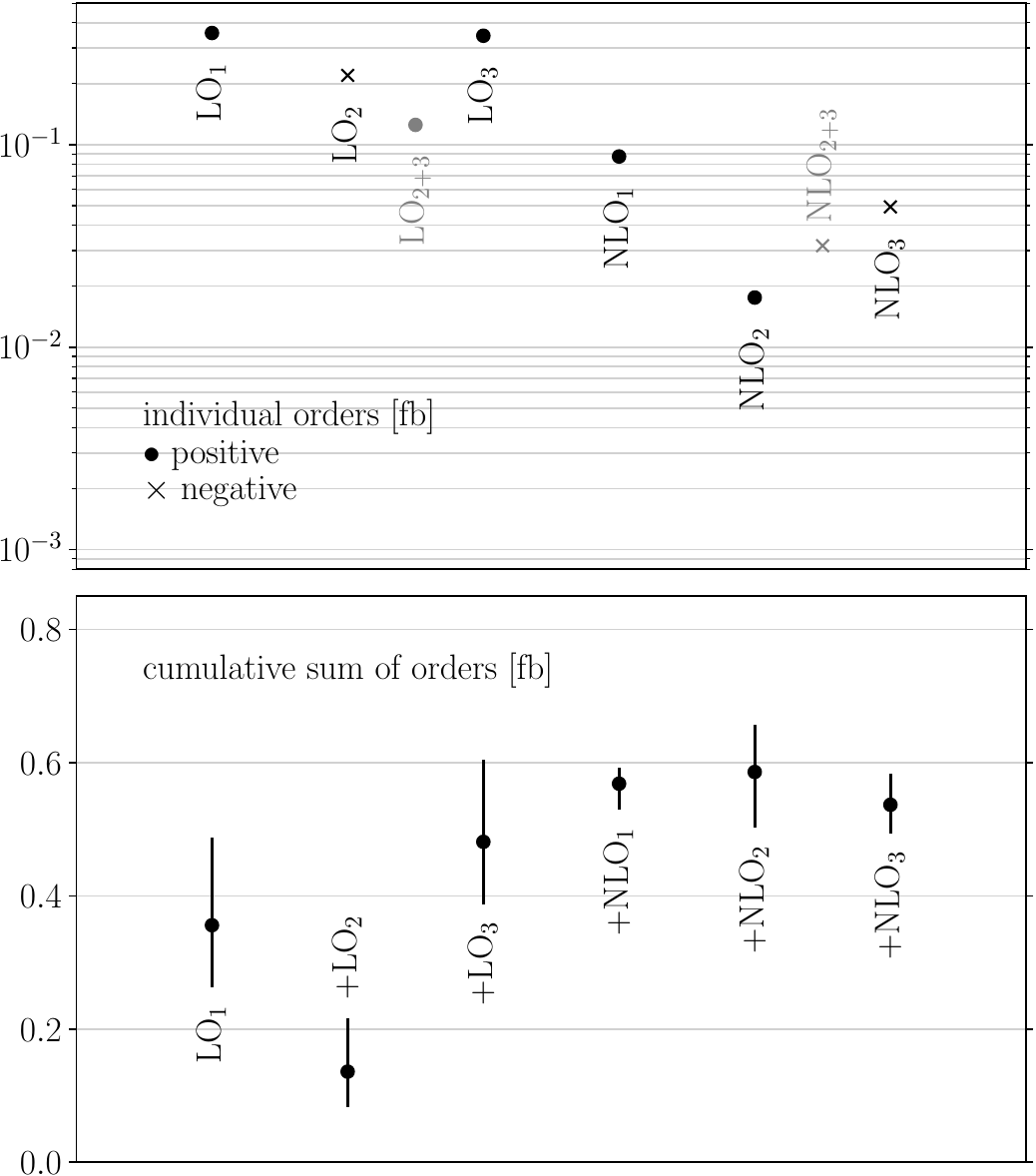}
\end{tabular}
\caption{Decomposition of the total cross section [fb] for $pp\to \bar{t}t\bar{t}W^{+}$ at $\sqrt{s}=13\,\mathrm{TeV}$, using the DR$_W^{40}$ prescription, without (left) and with (right) veto on extra $b$-jet radiation, which does not affect leading orders.
The top row shows individual (N)LO$_i$ contributions and the bottom row their cumulative sum taken from left to right.
Negative individual orders are marked with a cross instead of a dot.
The central values are evaluated at the reference scale $\mu_R=\mu_F=H_T/4$ and the scale uncertainties displayed are obtained from a nine-point variation.
}%
\label{fig:rates}%
\end{figure}

Let us now examine the complete set of electroweak LO and NLO orders.
In~\cref{tab:incl_xsec_lo_dr3} and \cref{fig:rates}, we report the total cross sections obtained using the $\mathrm{DR}_{W}^{40}$ prescription, with and without $b$-jet veto.
We include the $\mathrm{LO}_{1}$, $\mathrm{LO}_{2}$ and $\mathrm{LO}_{3}$ contributions and their corresponding NLO corrections.
Orders beyond (N)LO$_3$ only have percent-level effects on our predictions and are therefore neglected.\footnote{%
In the DR$_{1}$ scheme, the total LO$_{4}$ cross-section amounts to 0.0035$^{+9\%}_{-11\%}$~fb and the individual NLO$_{4}$ and NLO$_{5}$ corrections are 0.0033 and 0.0015~fb, respectively.}

\begin{table}[t]
\centering
\renewcommand{\arraystretch}{1.3}
\setlength{\tabcolsep}{5pt}
\newlength{\numsep}
\setlength{\numsep}{0.8em}
\newlength{\errsep}
\setlength{\errsep}{0.2em}

\resizebox{\textwidth}{!}{%
\begin{tabular}{c||c|c||c|c}
\toprule
\textbf{LO} &
\multicolumn{2}{c||}{\textbf{$\mathrm{DR}_{W}^{40}$}} &
\multicolumn{2}{c}{\textbf{$\mathrm{DR}_{W}^{40} + b\textrm{-veto}$}} \\
\cmidrule(r){1-1}
\cmidrule(lr){2-3}
\cmidrule(l){4-5}
\textbf{Predictions} &
\textbf{Corrections} & \textbf{Predictions} &
\textbf{Corrections} & \textbf{Predictions} \\
\midrule

\begin{tabular}[t]{@{}l@{\hspace{\numsep}}S[table-format=-1.3]@{\hspace{\errsep}}c@{}}
$\mathrm{LO~QCD}$ & 0.355  & {$^{+36\%}_{-26\%}$} \\
$\mathrm{LO}_{2}$            & -0.219 & {$^{+19\%}_{-23\%}$} \\
$\mathrm{LO}_{3}$            & 0.345  & {$^{+13\%}_{-12\%}$} \\
\addlinespace[3pt]
$\mathrm{LO}$        &  0.481 & {$^{+25\%}_{-19\%}$}
\end{tabular}
&
\begin{tabular}[t]{@{}l@{\hspace{\numsep}}S[table-format=-1.3]@{}}
$\mathrm{NLO}_{1}$ & 0.228 \\
$\mathrm{NLO}_{2}$ & -0.045 \\
$\mathrm{NLO}_{3}$ & 0.044 \\
$\mathrm{NLO}_{1+2+3}$ & 0.227
\end{tabular}
&
\begin{tabular}[t]{@{}l@{\hspace{\numsep}}S[table-format=-1.3]@{\hspace{\errsep}}c@{}}
$\mathrm{NLO~QCD}$       & 0.583 & {$^{+18\%}_{-16\%}$} \\
\addlinespace[3pt]
$\mathrm{LO}+\mathrm{NLO}_1$        & 0.710 & {$^{+14\%}_{-13\%}$}\\
$\mathrm{NLO}$                      & 0.708 & {$^{+15\%}_{-12\%}$}
\end{tabular}
&
\begin{tabular}[t]{@{}l@{\hspace{\numsep}}S[table-format=-1.3]@{}}
$\mathrm{NLO}_{1}$ & 0.087 \\
$\mathrm{NLO}_{2}$ & 0.017 \\
$\mathrm{NLO}_{3}$ & -0.049 \\
$\mathrm{NLO}_{1+2+3}$ & 0.055
\end{tabular}
&
\begin{tabular}[t]{@{}l@{\hspace{\numsep}}S[table-format=-1.3]@{\hspace{\errsep}}c@{}}
$\mathrm{NLO~QCD}$       & 0.442 & {$^{+9\%}_{-10\%}$} \\
\addlinespace[3pt]
$\mathrm{LO}+\mathrm{NLO}_1$        & 0.569 & {$^{+4\%}_{-7\%}$}\\
$\mathrm{NLO}$                      & 0.536 & {$^{+8\%}_{-8\%}$}
\end{tabular}
\\
\bottomrule
\end{tabular}%
}
\caption{%
LO and NLO cross sections [fb] for $\bar t t \bar t W^+$ production at the 13~TeV LHC, using the $\mathrm{DR}_{W}^{40}$ prescription, without and with veto on extra $b$-jet radiation.
The first column shows the LO contributions, whilst the remaining columns show the individual NLO corrections and full-fledged predictions.
Scale uncertainties are obtained from the envelope of QCD nine-point $(\mu_R,\mu_F)$ variations, with $\mu_{R,F}\in\{0.5,1,2\}\mu_0$ and $\mu_0=H_T/4$.
}
\label{tab:incl_xsec_lo_dr3}
\end{table}

At leading order, LO$_2$ and LO$_3$ have opposite signs but only partially cancel each other.
Together, they increase the LO$_1$ inclusive rate by 35\%.
They should thus be included.
At next-to-leading order, the NLO$_2$ and NLO$_3$ contributions also have opposite signs.
Without $b$-jet veto, they cancel to a very high degree: their sum is more than an order of magnitude smaller than their individual absolute values (see top left panel of \cref{fig:rates}).
Although the individual NLO$_i$'s have a sizeable scale dependence (largely compensated by the corresponding LO$_i$ contributions), the cancellation between NLO$_2$ and NLO$_3$ is robust against scale variation.
As can be seen in the bottom left panel of \cref{fig:rates}, the inclusive prediction including orders up to NLO$_1$ (i.e.\ LO + NLO$_1$) is therefore essentially identical to the one also including NLO$_2$ and NLO$_3$ (i.e.\ NLO).
We will investigate in the next section whether such a remarkable cancellation, similar to the one observed in four-top production~\cite{Frederix:2017wme}, persists at the differential level.

With a $b$-jet veto (see right panels of \cref{fig:rates}), the cancellation between NLO$_2$ and NLO$_3$ is less significant (note their signs are flipped compared to the case without veto).
Nevertheless, the LO + NLO$_1$ prediction still only departs from the complete NLO one by about 6\%.
Moreover, the scale-variation ranges of these two predictions largely overlap.

\subsection{Differential distributions}

\begin{figure}[p]\centering
\vspace*{-5mm}%
\begin{tabular}{@{}c@{\:}c@{}}
\qquad DR$_W^{40}$
&\qquad DR$_W^{40}$ + $b$-veto
\\
\includegraphics[width=.5\textwidth]{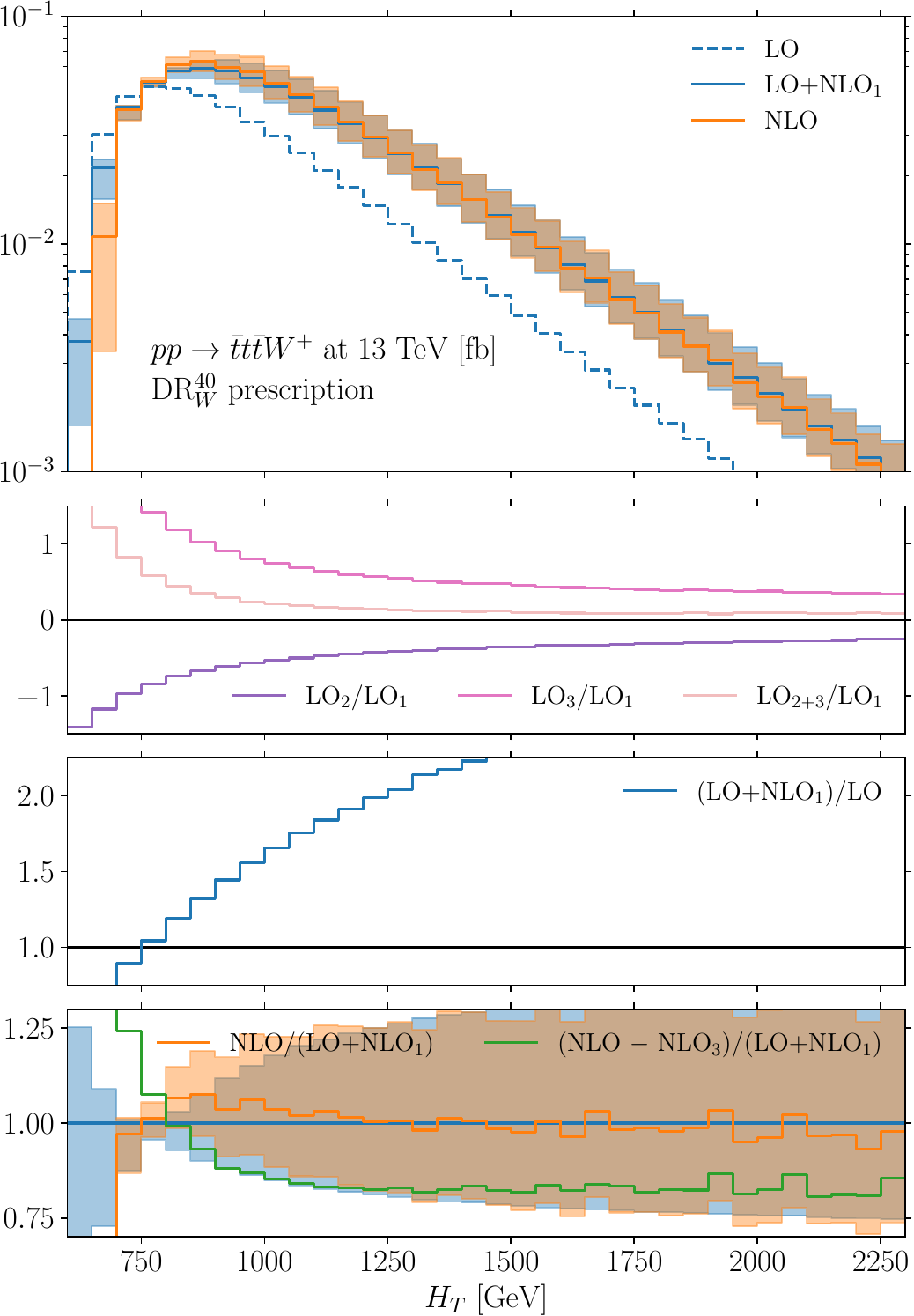}%
&\includegraphics[width=.5\textwidth]{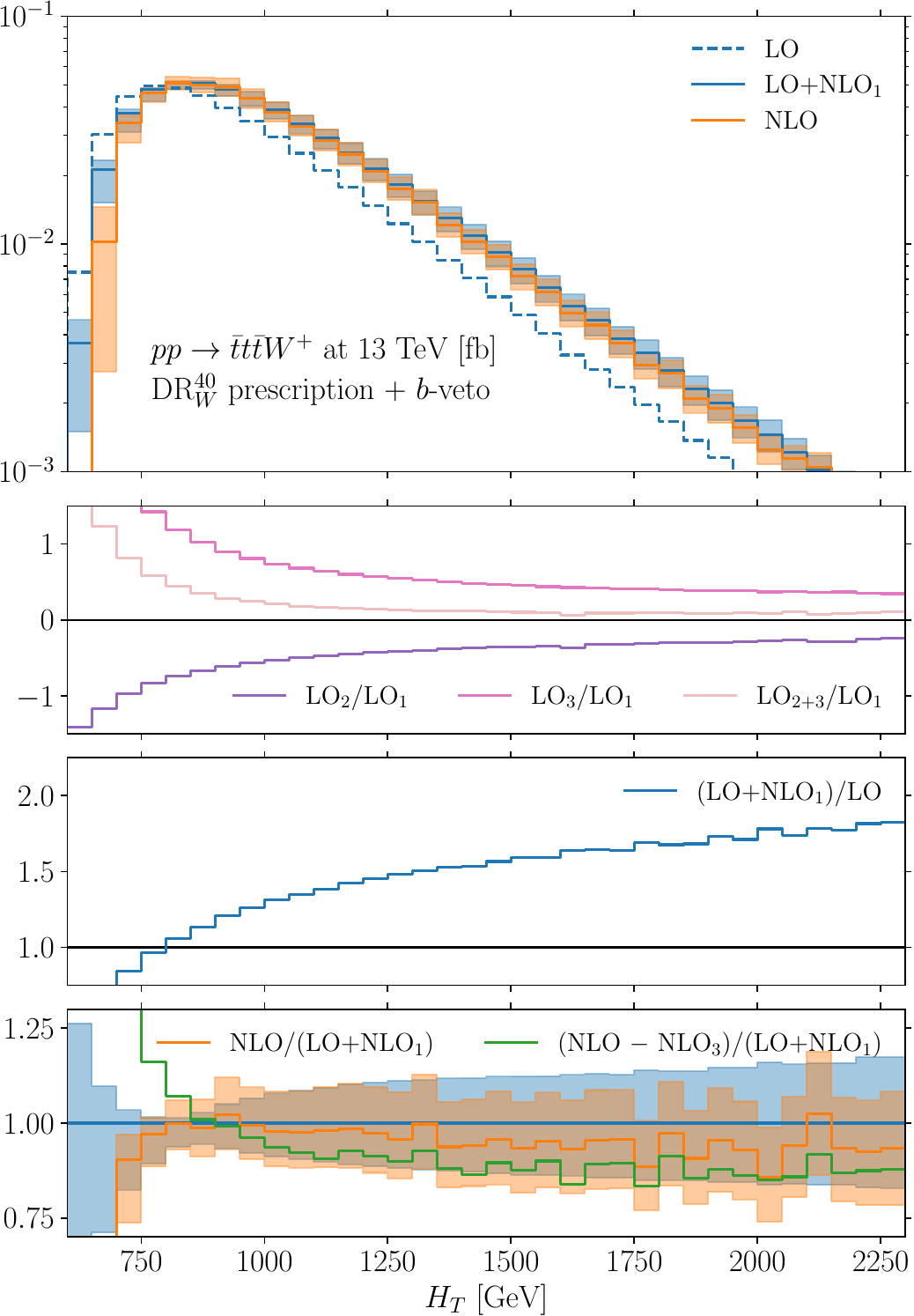}%
\\[4mm]
\includegraphics[width=.5\textwidth]{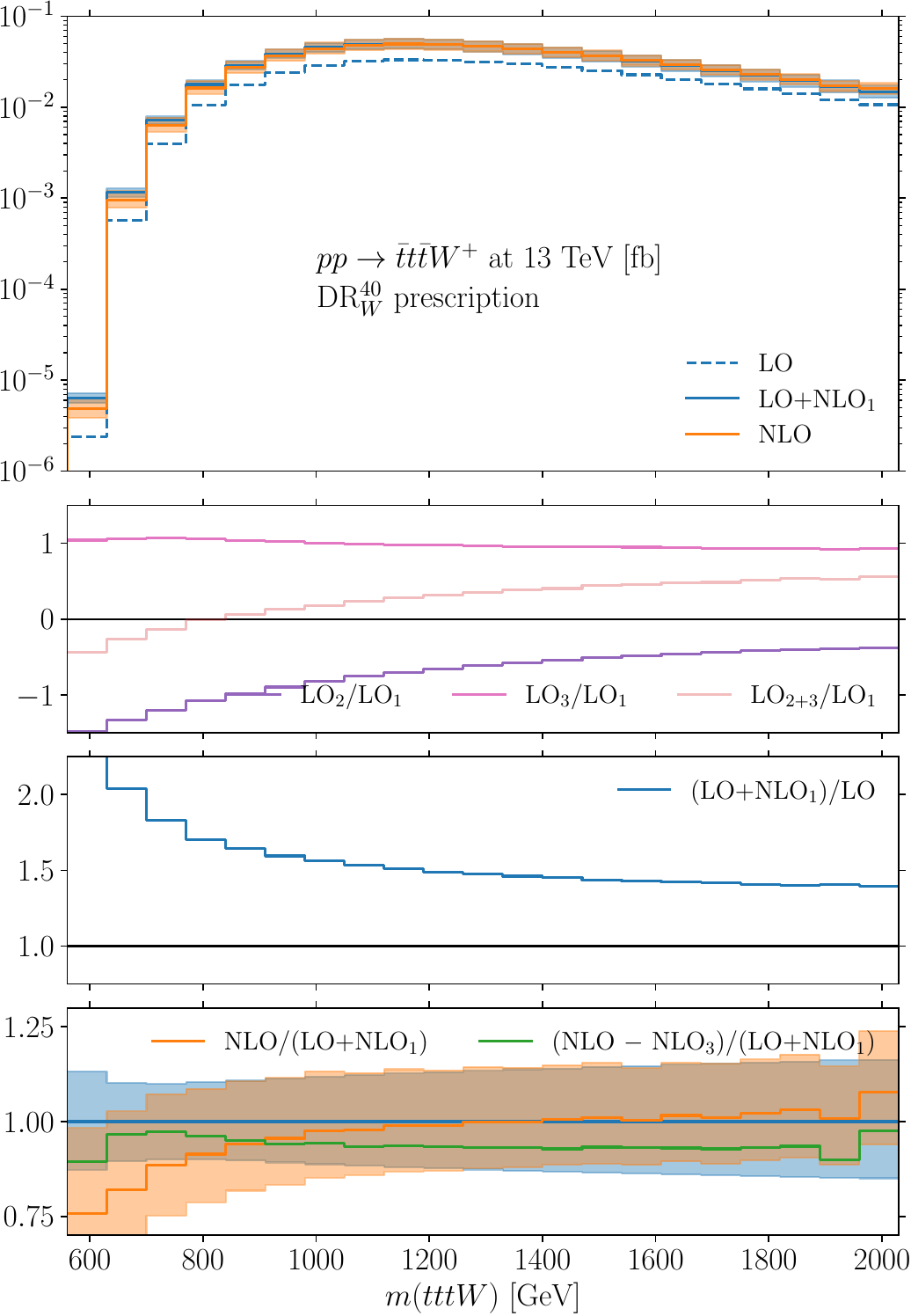}%
&\includegraphics[width=.5\textwidth]{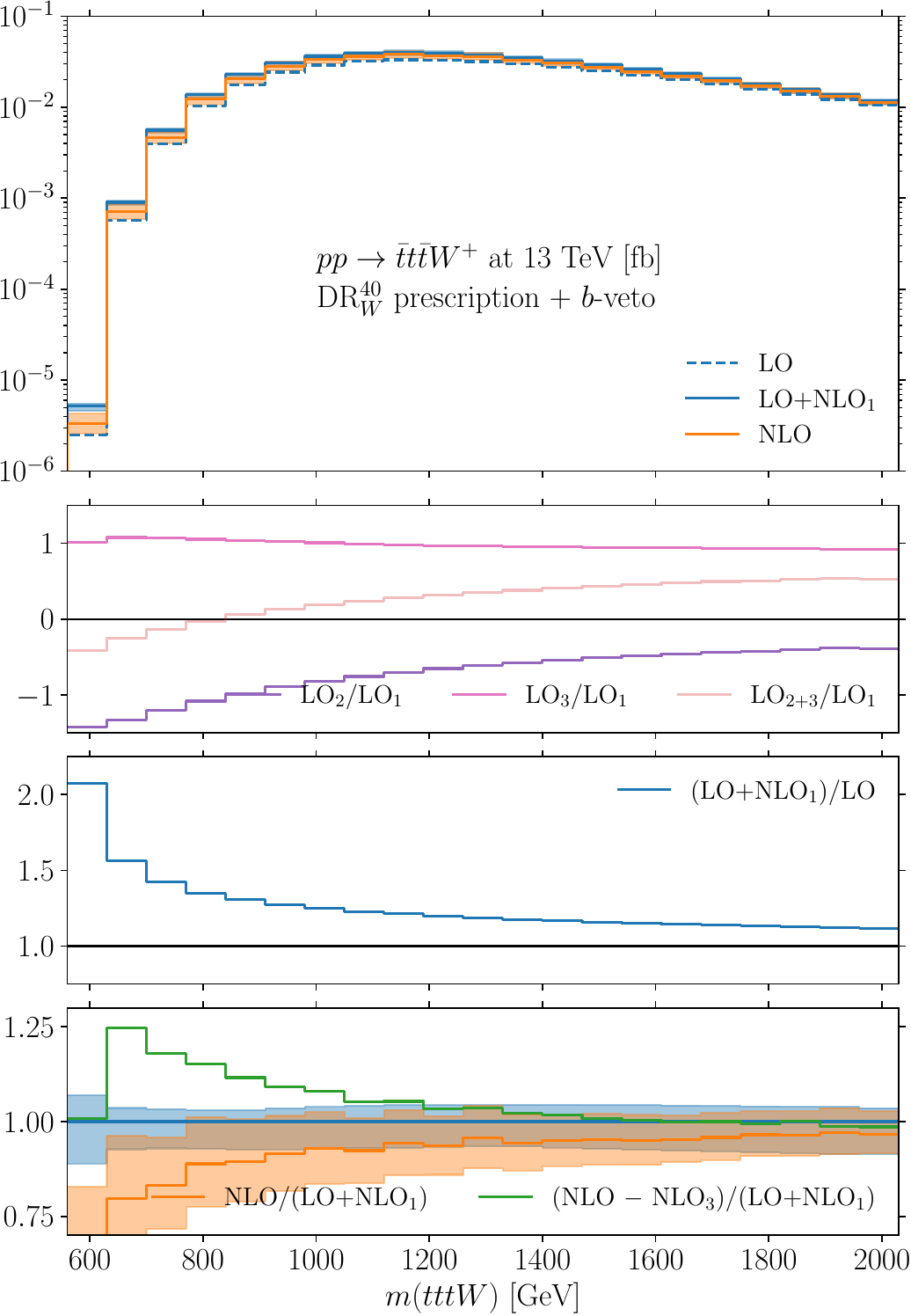}%
\end{tabular}
\caption{Differential distributions of the scalar sum of transverse energies ($H_T$) and $tttW$ invariant mass in $pp\to\ttop$ production at 13~TeV, detailing the various leading and next-to-leading order contributions.
}%
\label{fig:diff-energies}%
\end{figure}%

\begin{figure}[p]\centering
\vspace*{-5mm}%
\begin{tabular}{@{}c@{\:}c@{}}
\qquad DR$_W^{40}$
&\qquad DR$_W^{40}$ + $b$-veto
\\
\includegraphics[width=.5\textwidth]{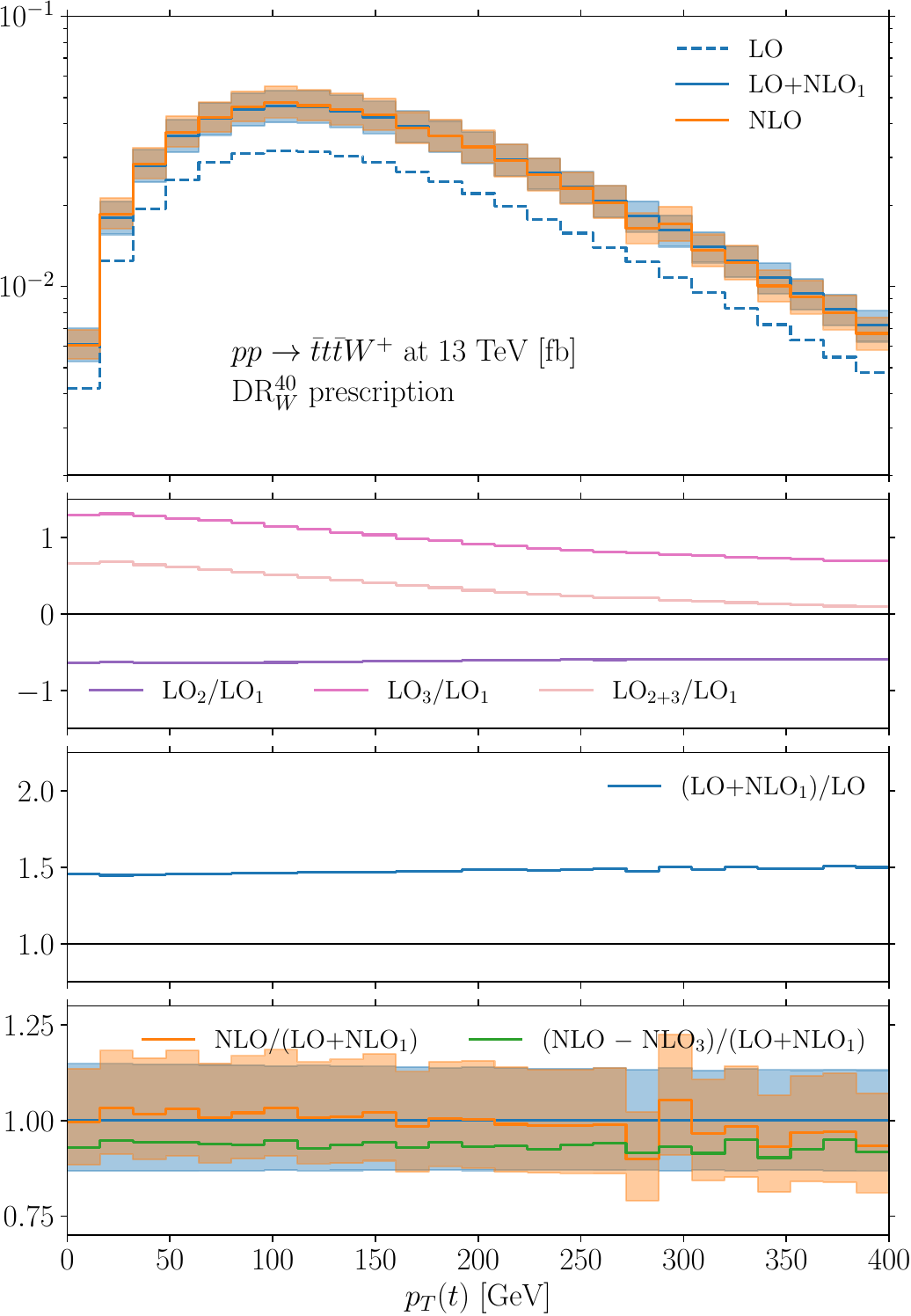}%
&\includegraphics[width=.5\textwidth]{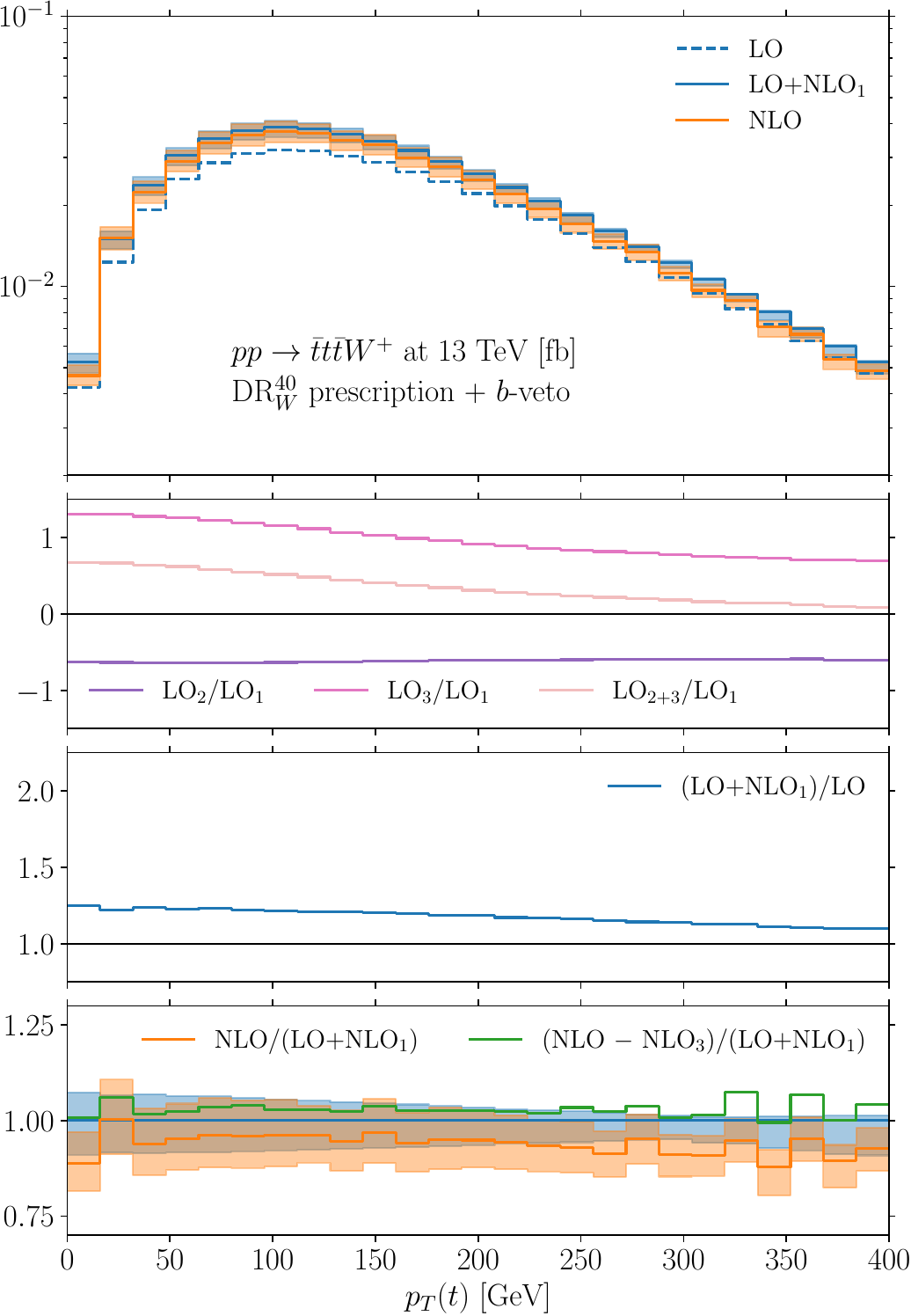}%
\\[4mm]
\includegraphics[width=.5\textwidth]{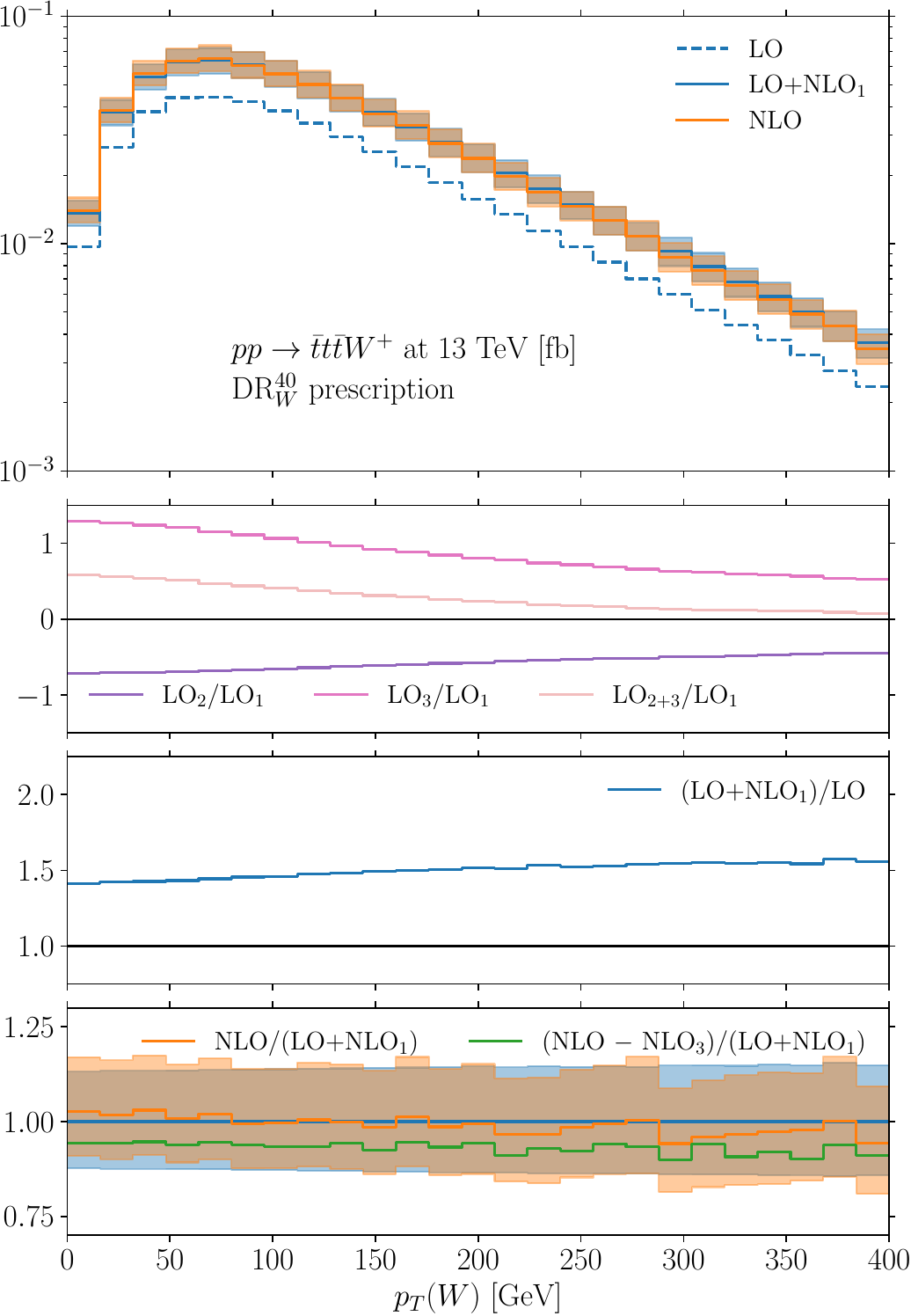}%
&\includegraphics[width=.5\textwidth]{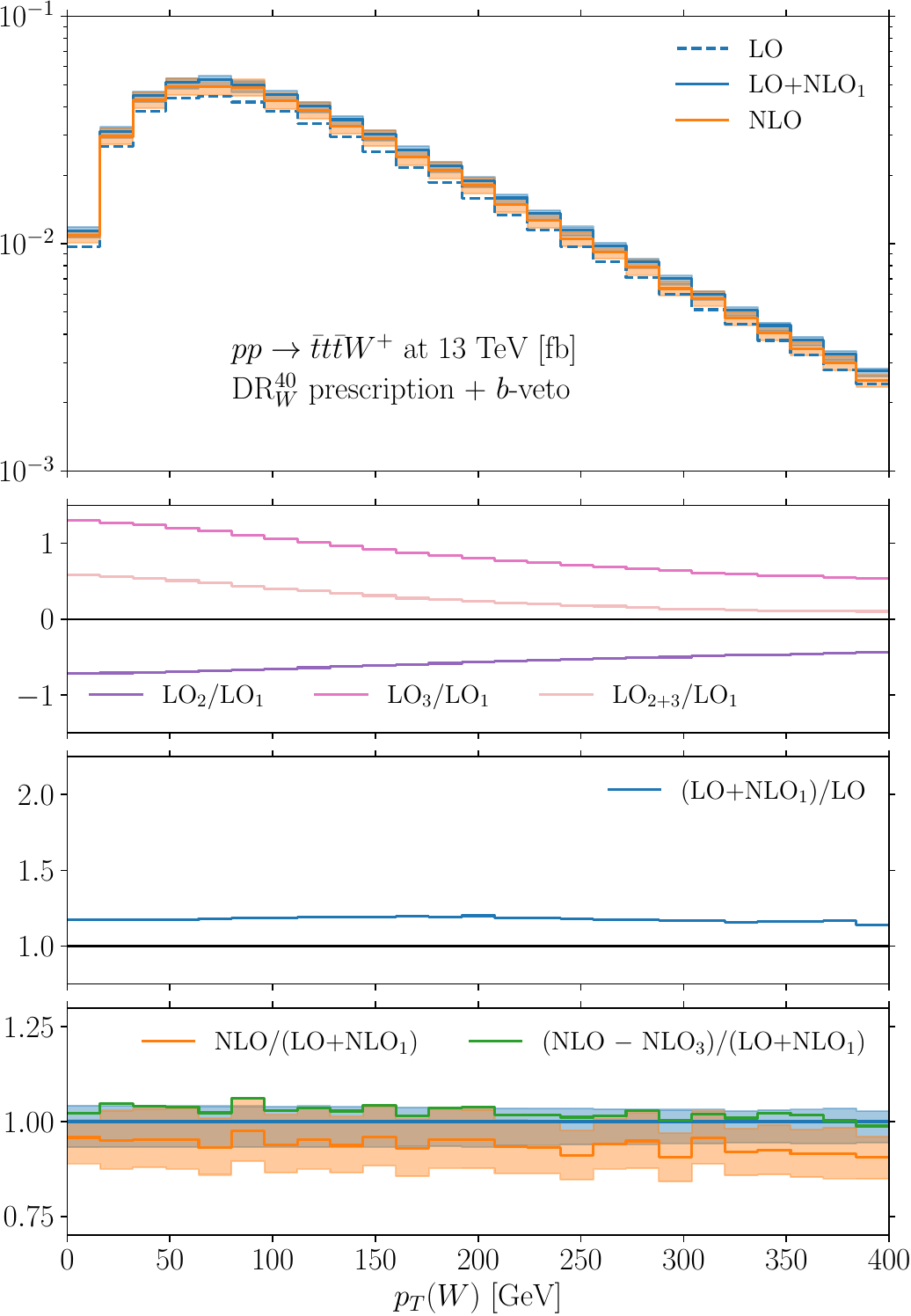}%
\end{tabular}
\caption{Transverse momentum distributions of the (positively charged) top quark and $W^+$ boson in $pp\to\ttop$ production at 13~TeV, detailing the various leading and next-to-leading order contributions.
}%
\label{fig:diff-pT}%
\end{figure}%

\begin{figure}[p]\centering
\vspace*{-5mm}%
\begin{tabular}{@{}c@{\:}c@{}}
\qquad DR$_W^{40}$
&\qquad DR$_W^{40}$ + $b$-veto
\\
\includegraphics[width=.5\textwidth]{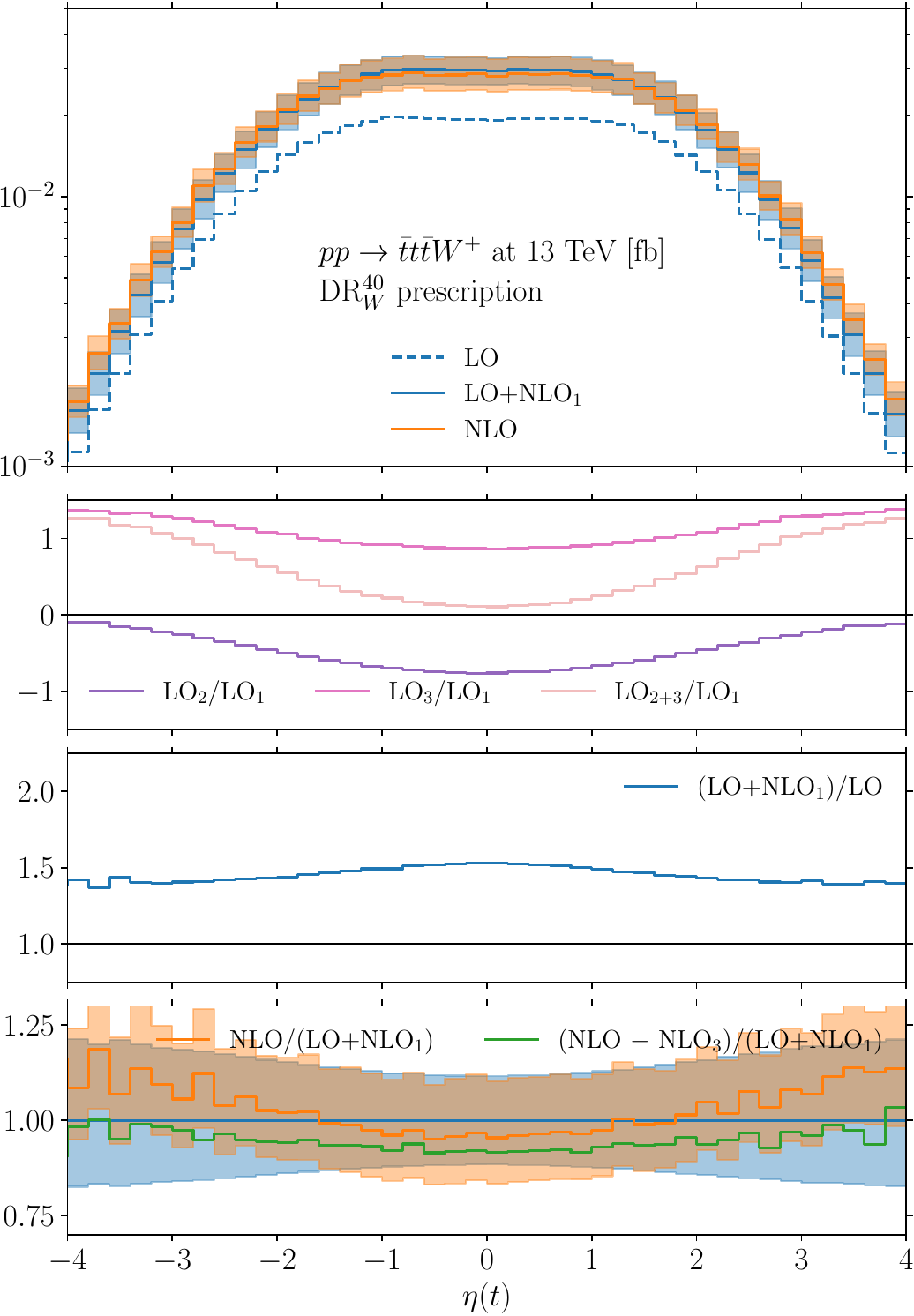}%
&\includegraphics[width=.5\textwidth]{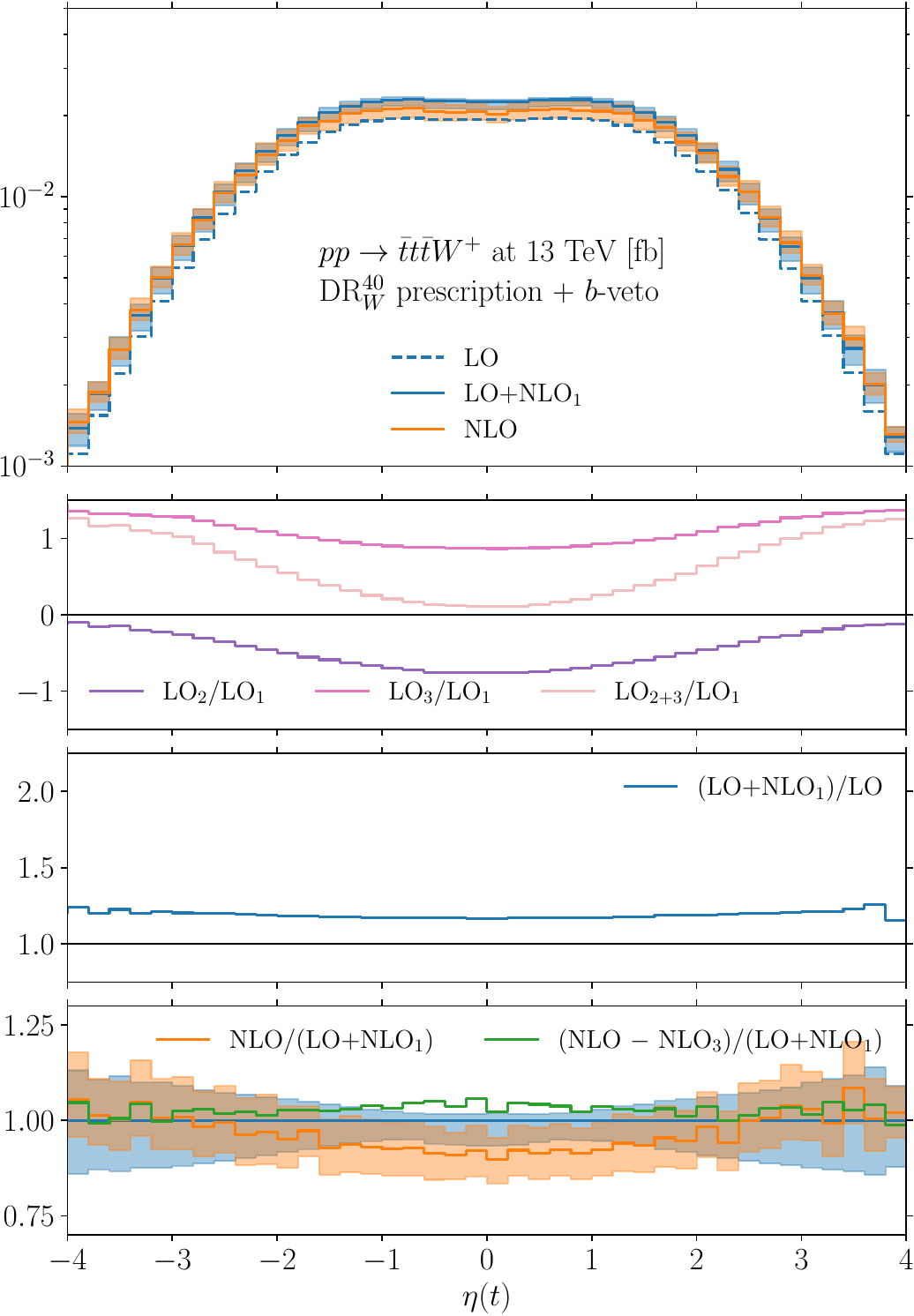}%
\\[4mm]
\includegraphics[width=.5\textwidth]{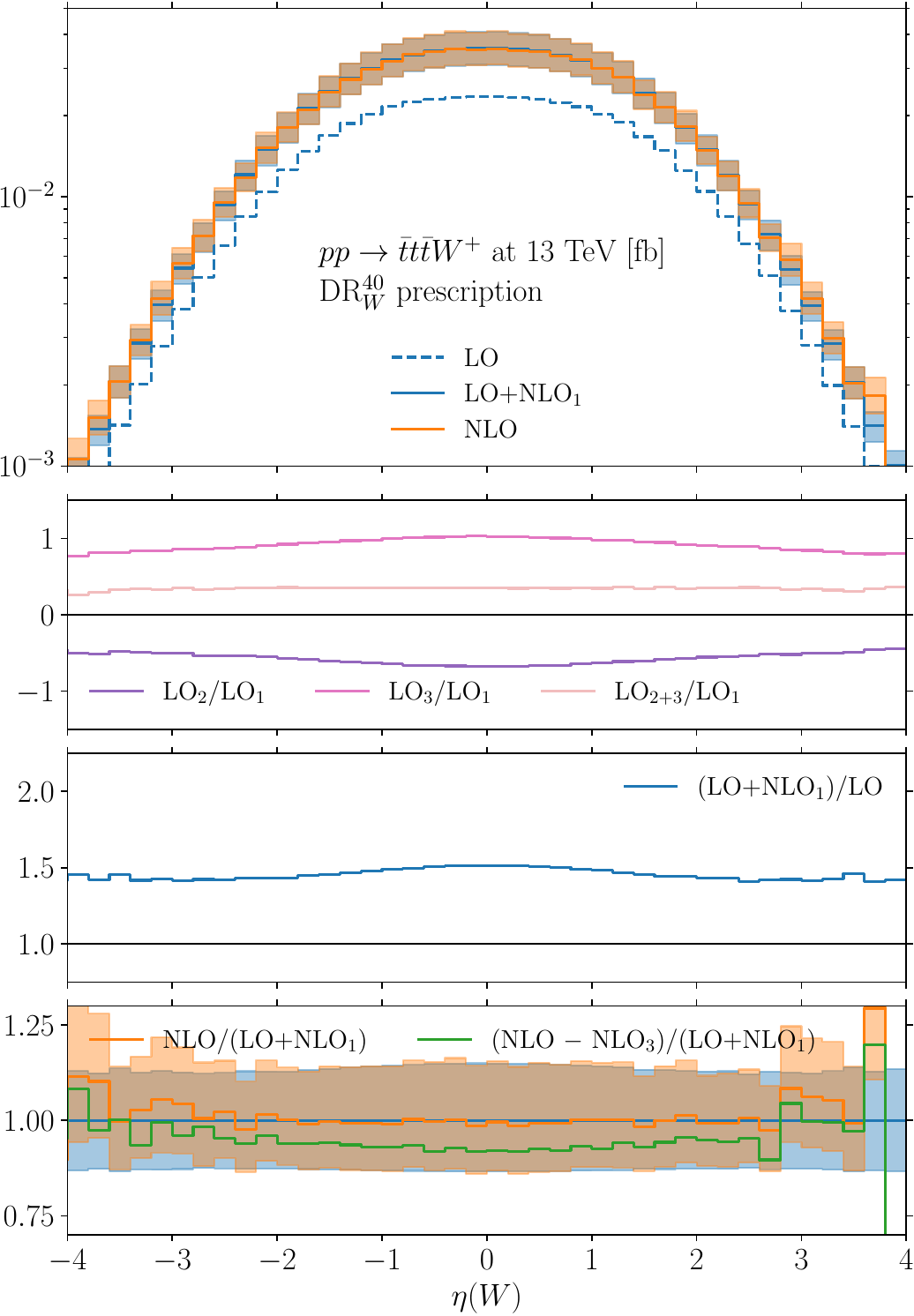}%
&\includegraphics[width=.5\textwidth]{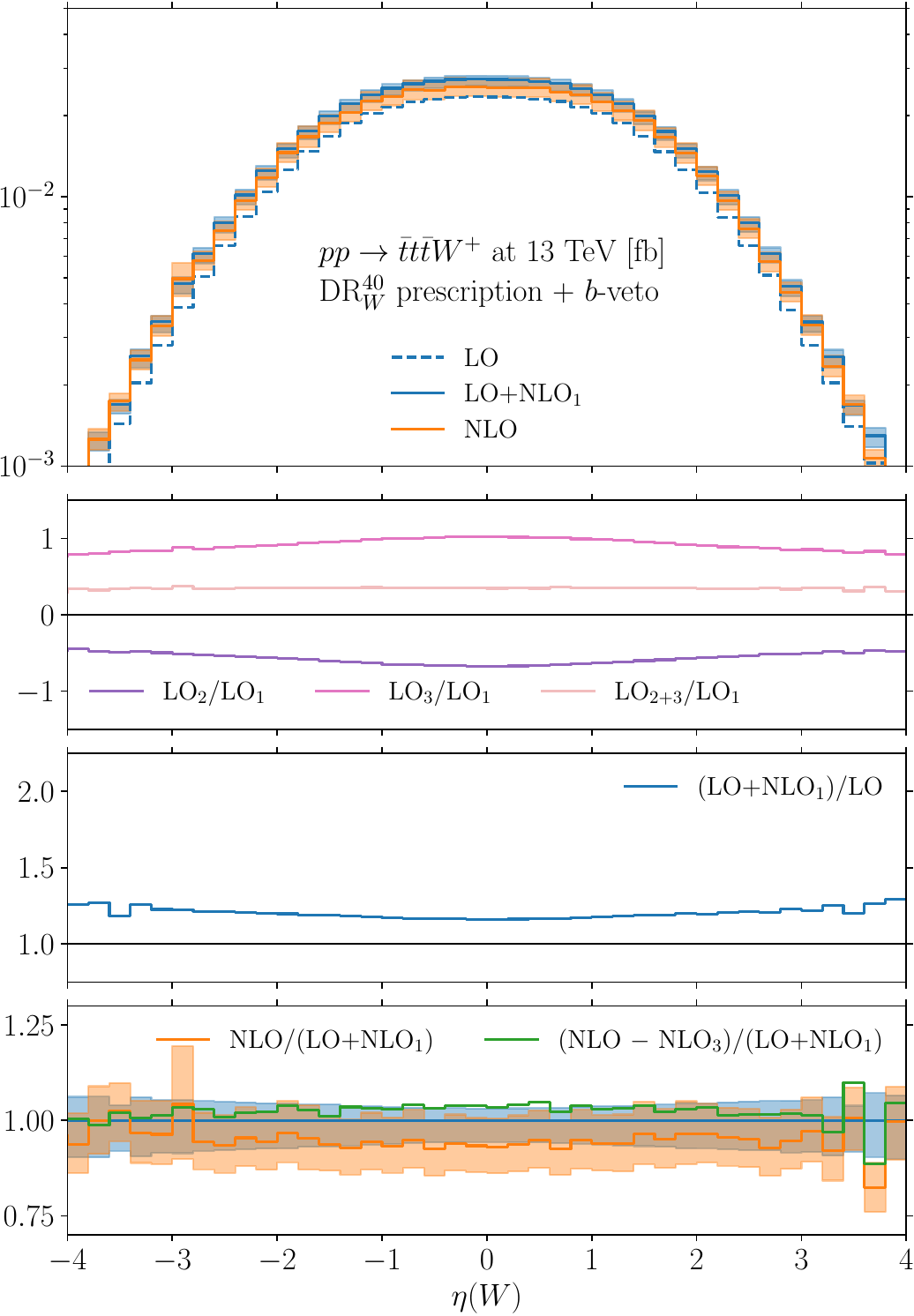}%
\end{tabular}
\caption{Pseudo-rapidity distributions of the (positively charged) top quark and $W^+$ boson in $pp\to\ttop$ production at 13~TeV, detailing the various leading and next-to-leading order contributions.
}%
\label{fig:diff-eta}%
\end{figure}%

We now examine the differential distributions of various kinematic variables: $H_T$ and $m(tttW)$ in \cref{fig:diff-energies}, $p_T(t)$ and $p_T(W)$ in \cref{fig:diff-pT}, $\eta(t)$ and $\eta(W)$ in \cref{fig:diff-eta}.
As in earlier figures, the left columns show the DR$_W^{40}$ predictions without the $b$-jet veto, while the right columns include the latter.
The various insets detail leading and next-to-leading order contributions.

From the first insets, one observes that the LO$_2$ and LO$_3$ contributions remain essentially of $\mathcal{O}(1)$ compared to LO$_1$ across the various distributions, without change of sign.
The LO$_3$/LO$_1$ ratio remains almost exactly equal to $1$ across the $m(tttW)$ distributions, while the LO$_2$/LO$_1$ one is particularly flat in the $p_T(t)$ spectrum.
The most noticeable variations occur across the $H_T$ distribution where LO$_2$ and LO$_3$ contributions are enhanced at threshold and progressively decay towards the high-energy tail.
Elsewhere, the degree of cancellation between LO$_2$ and LO$_3$ contributions also remains rather constant.

The second insets display the relative impact of NLO$_1$ corrections.
This $K$-factor is relatively stable across distributions, close to 1.5 in the absence of $b$-veto and to 1.25 when a $b$-veto is applied.
`Giant' $K$-factors are however visible in the $H_T$ distribution, a well-known feature (see e.g.\ \cite{Frixione:1992pj, Maltoni:2015ena, Rubin:2010xp}).
Elsewhere, the $m(tttW)$ invariant mass is the distribution showing the most significant differential variation, with a clear enhancement of the $K$-factor towards the threshold.
The absolute rate in this region is however much suppressed.

The third insets focus on the impact of NLO$_2$ and NLO$_3$ corrections.
Since each of these presents a large scale dependence, we only show them summed with lower order corrections.
Histograms are normalised, in these insets, by the central (LO + NLO$_1$) prediction, for which the nine-point scale variation is also displayed as a blue band.
The main information is contained in the orange NLO/(LO + NLO$_1$) ratio, also shown with its scale variation.
It remains close to unity across most differential distributions, showing that the NLO$_1$ corrections do account for the bulk of NLO effects.
The formally subleading NLO$_{2}$ and NLO$_3$ corrections are enhanced near threshold in the $m(tttW)$ invariant mass distributions.
They can reach tens of percent, which is of the order of the scale variation.
In this region, the absolute rate is however very suppressed and resummation would become necessary.

Still considering the last insets of figs.~\ref{fig:diff-energies}, \ref{fig:diff-pT}, \ref{fig:diff-eta}, let us focus now first on the predictions without $b$-jet veto (left column).
The green (NLO $-$ NLO$_3$)/(LO + NLO$_1$) histograms, displayed without scale variation to avoid clutter, show that the NLO$_2$ contributions are negative essentially everywhere across the phase space.
The NLO$_3$ corrections, included in the orange NLO/(LO + NLO$_1$) histograms, bring these back up and closer to unity.
Remarkably, this shows that the high degree of cancellation between NLO$_{2}$ and NLO$_3$ corrections observed at the inclusive level is also largely maintained differentially.
The introduction of a $b$-jet veto (right column) flips the signs of these NLO$_{2}$ and NLO$_3$ corrections and partially spoils their cancellation.
The net downward effect of about 6\% observed at the inclusive level does not show a significant differential dependence.
It also remains everywhere of the same order as the scale variation.

\section{Summary and conclusions}
\label{sec:concl}

Motivated by its contamination of experimental selections targeting four top-quark production ($tttt$), we investigate the production of three top quarks at NLO.
We focus on the associated production with a $W$ boson ($tttW$) which constitutes the largest three-top-production mode.
The $tttW$ process is well-defined at LO and in the five-flavour scheme where $b$ quarks are massless and included in the parton distribution function of the proton.
At NLO however, the radiation of a $b$ jet produces the same $tttWb$ final state as $tttt$ production with a $t\to bW$ decay.
The $tttW$ and $tttt$ processes thus become inseparable.
Attempting to treat the diagrams of both types in isolation breaks gauge invariance and leads to sizeable unitarity violation away from the on-shell $m({Wb})\to m_t$ limit.
This effect is also present when attempting to separate $tW$ diagrams from top-quark-pair production ones, but is less significant in relative terms.

For general kinematics, only joint $tttW+tttt$ predictions are therefore meaningful.
Splitting this computation in two components is nevertheless the most efficient way of achieving NLO accuracy in both on-shell and off-shell kinematic regimes.
The on-shell $tttt$ production is well-defined and the invariant masses of the top decay products are commonly smeared away from the strict on-shell limit.
We propose to restrict this prediction to a relatively narrow on-shell window around $m({Wb})=m_t$ to constitute the first component of the joint $tttW+tttt$ prediction.
The second component is then defined to include all the remaining contributions.
It thus only excludes the square of $tttt$ diagrams inside the on-shell window.
There, it therefore accounts for $tttW$ diagrams squared and for their interference with $tttt$ ones.
Outside of the on-shell window, both resonant $tttt$ and non-resonant $tttW$ diagrams are included, ensuring gauge invariance and avoiding unitarity violation in energy distributions.
An on-shell window width of $\pm 40$~GeV is found to be wide enough to describe most of the on-shell region with the higher-order on-shell $tttt$ prediction, while remaining narrow enough to limit the residual gauge dependence that could survive inside the window.
The simple and well-behaved prescription described above is a variant of the so-called diagram removal procedures.
We therefore denote it by DR$_W^{40}$, where the subscript `$W$' stands for `window' and the superscript `$40$' specifies the on-shell window half-width in GeV.

Beside the joint $tttW+tttt$ prediction obtained following such a prescription which is consistent for general kinematics, we also consider imposing an idealised veto on extra $b$-jet radiation in $tttW$ production.
It efficiently suppresses both $tttt$ diagrams and their interference with $tttW$ ones, providing a restriction of the phase space where $tttW$ diagrams alone become approximately well-defined.
Unlike in simpler processes like $tW$ production where a similar veto can be enforced on data, there are probably too many $b$ jets produced by the top-quark decays to realistically apply it in $tttW$ production.
Nevertheless, such a vetoed prediction can serve as target for the definition of a realistic $tttW$-enriched selection, where data can still be compared to the joint $tttW+tttt$ prediction unless a sufficient purity is achieved.
Quantum-information observables based on top-quark spin correlations were recently found to distinguish LO $tttj$ and $tttt$ samples~\cite{Durupt:2025wuk} and could potentially be involved in the selection of three-top-like events.
The obtained $tttW$-enriched selection would allow for a more granular test of the SM in multi-top-quark production.
Alternatively, more specific BSM hypotheses could be tested directly, using simplified models or the SM effective field theory (SMEFT).
These would not require the artificial separation between classes of diagrams (BSM would often actually contribute to both of them) as sensitivity could be optimised directly against a consistent joint $tttW+tttt$ prediction in the SM.

Equipped with such simple and well-behaved prescriptions, we finally compute complete NLO predictions for $tttW$ production.
Studying the scale dependence at NLO QCD, we find that central dynamical scales of $H_T/4$ lead to a reasonable perturbative expansion, with moderate $K$-factor and scale uncertainties.
As observed in $tttt$ previously, several formally subleading electroweak contributions are actually numerically significant.
Some of them however efficiently cancel against one another, both at the inclusive and differential levels.
As a result, a prediction including the first three LO's and the NLO QCD correction provides a reasonable approximation for the complete NLO prediction.
Conveniently, it can be matched to parton shower to provide realistic fully differential event samples that can be passed through detector simulation.
By reducing the LO $tttt$ contamination, the $b$ veto further reduces both the $K$-factor and scale uncertainty.
It also partially spoils the cancellation between subleading electroweak orders.

Our predictions close a gap in the theoretical description of multi-top production at the LHC.
Three-top production, though rare, plays a non-negligible role as an uncertainty source in four-top measurements and in searches for new physics that couple preferentially to top quarks.
By providing complete NLO accuracy, we enable more robust interpretations of experimental data and improve the reliability of background estimates in analyses targeting e.g.\ top-philic BSM physics.
Since the current four-top experimental analyses cannot efficiently isolate the on-shell $tttt$ region, we argue that they should be compared to a joint $tttW+tttt$ prediction.
Our DR$_W^{40}$ computation indicates that its rate, counting only $t\bar{t}tW^{-}$ and $\bar{t}t\bar{t}W^{+}$ conjugate processes, is more than $1.4\:$fb higher than the on-shell $tttt$ one.
This is a sizeable correction of more than $10\%$.

Directions for future work include the study of the $tttj$ production mode and of its overlap with $tttW$.
Given the smaller hierarchy between their respective rates, we however expect less significant issues than with the $tttW$--$tttt$ overlap considered here.
The matching of our fixed-order results to parton showers and the NLO description of top-quark decays could also be performed.
Exploring the interplay between three and four top-quark production in probing SMEFT operators, or explicit BSM scenarios, is another natural extension of the present work.
In particular, deformations of the SM could affect the cancellation between sizeable electroweak orders and yield enhanced sensitivity.

\acknowledgments
We are grateful to Didar Dobur, Valentin Hirschi, Markus Luty, Fabio Maltoni and David Marckx for valuable discussions and, in particular, to Marcel Vos for having triggered the first computations.
G.D.~is a Research Associate of the Fund for Scientific Research~--~FNRS, Belgium, supported in part by the IISN convention 4.4509.26.
H.F.~acknowledges support by the European Research Council (ERC) under the European Union's Horizon 2020 research and innovation programme (Grant agreement No.~949451) and by a Royal Society University Research Fellowship through grant URF/R1/201553.
H.F.~is supported by the European Union under the MSCA fellowship (Grant agreement No.~101208909).
D.P.~and M.Z.~acknowledge the financial support by the MUR (Italy), with funds of the European Union (NextGenerationEU), through the PRIN2022 grant 2022EZ3S3F.
Computational resources have been provided by the supercomputing facilities of the Université catholique de Louvain (CISM/UCL) and the Consortium des Équipements de Calcul Intensif en Fédération Wallonie Bruxelles (CÉCI) funded by the Fonds de la Recherche Scientifique de Belgique (F.R.S.-FNRS) under convention 2.5020.11 and by the Walloon Region.

\end{fmffile}
\bibliographystyle{JHEP}
\bibliography{bibliography}

\providecommand{\href}[2]{#2}\begingroup\raggedright\begin{thebibliography}{10}

\bibitem{Cao:2016wib}
Q.-H.~Cao, S.-L.~Chen and Y.~Liu, \emph{{Probing Higgs Width and Top Quark
  Yukawa Coupling from $t\bar{t}H$ and $t\bar{t}t\bar{t}$ Productions}},
  \href{https://doi.org/10.1103/PhysRevD.95.053004}{\emph{Phys. Rev. D}
  {\bfseries 95} (2017) 053004}
  [\href{https://arxiv.org/abs/1602.01934}{{\ttfamily 1602.01934}}].

\bibitem{Cao:2019ygh}
Q.-H.~Cao, S.-L.~Chen, Y.~Liu, R.~Zhang and Y.~Zhang, \emph{{Limiting top
  quark-Higgs boson interaction and Higgs-boson width from multitop
  productions}}, \href{https://doi.org/10.1103/PhysRevD.99.113003}{\emph{Phys.
  Rev. D} {\bfseries 99} (2019) 113003}
  [\href{https://arxiv.org/abs/1901.04567}{{\ttfamily 1901.04567}}].

\bibitem{ATLAS:2024mhs}
{\scshape ATLAS} collaboration, \emph{{Constraint on the total width of the
  Higgs boson from Higgs boson and four-top-quark measurements in pp collisions
  at s = 13 TeV with the ATLAS detector}},
  \href{https://doi.org/10.1016/j.physletb.2025.139277}{\emph{Phys. Lett. B}
  {\bfseries 861} (2025) 139277}
  [\href{https://arxiv.org/abs/2407.10631}{{\ttfamily 2407.10631}}].

\bibitem{Zhang:2017mls}
C.~Zhang, \emph{{Constraining $qqtt$ operators from four-top production: a case
  for enhanced EFT sensitivity}},
  \href{https://doi.org/10.1088/1674-1137/42/2/023104}{\emph{Chin. Phys. C}
  {\bfseries 42} (2018) 023104}
  [\href{https://arxiv.org/abs/1708.05928}{{\ttfamily 1708.05928}}].

\bibitem{Malekhosseini:2018fgp}
M.~Malekhosseini, M.~Ghominejad, H.~Khanpour and M.~Mohammadi~Najafabadi,
  \emph{{Constraining top quark flavor violation and dipole moments through
  three and four-top quark productions at the LHC}},
  \href{https://doi.org/10.1103/PhysRevD.98.095001}{\emph{Phys. Rev. D}
  {\bfseries 98} (2018) 095001}
  [\href{https://arxiv.org/abs/1804.05598}{{\ttfamily 1804.05598}}].

\bibitem{Darme:2018dvz}
L.~Darm{\'e}, B.~Fuks and M.~Goodsell, \emph{{Cornering sgluons with
  four-top-quark events}},
  \href{https://doi.org/10.1016/j.physletb.2018.08.001}{\emph{Phys. Lett. B}
  {\bfseries 784} (2018) 223}
  [\href{https://arxiv.org/abs/1805.10835}{{\ttfamily 1805.10835}}].

\bibitem{Hou:2019gpn}
W.-S.~Hou, M.~Kohda and T.~Modak, \emph{{Implications of Four-Top and Top-Pair
  Studies on Triple-Top Production}},
  \href{https://doi.org/10.1016/j.physletb.2019.134953}{\emph{Phys. Lett. B}
  {\bfseries 798} (2019) 134953}
  [\href{https://arxiv.org/abs/1906.09703}{{\ttfamily 1906.09703}}].

\bibitem{Cao:2019qrb}
Q.-H.~Cao, S.-L.~Chen, Y.~Liu and X.-P.~Wang, \emph{{What can We Learn from
  Triple Top-Quark Production?}},
  \href{https://doi.org/10.1103/PhysRevD.100.055035}{\emph{Phys. Rev. D}
  {\bfseries 100} (2019) 055035}
  [\href{https://arxiv.org/abs/1901.04643}{{\ttfamily 1901.04643}}].

\bibitem{Alvarez:2019uxp}
E.~Alvarez, A.~Juste and R.M.S.~Seoane, \emph{{Four-top as probe of light
  top-philic New Physics}},
  \href{https://doi.org/10.1007/JHEP12(2019)080}{\emph{JHEP} {\bfseries 12}
  (2019) 080} [\href{https://arxiv.org/abs/1910.09581}{{\ttfamily
  1910.09581}}].

\bibitem{Banelli:2020iau}
G.~Banelli, E.~Salvioni, J.~Serra, T.~Theil and A.~Weiler, \emph{{The Present
  and Future of Four Top Operators}},
  \href{https://doi.org/10.1007/JHEP02(2021)043}{\emph{JHEP} {\bfseries 02}
  (2021) 043} [\href{https://arxiv.org/abs/2010.05915}{{\ttfamily
  2010.05915}}].

\bibitem{Hou:2020chc}
W.-S.~Hou and T.~Modak, \emph{{Probing Top Changing Neutral Higgs Couplings at
  Colliders}}, \href{https://doi.org/10.1142/S0217732321300068}{\emph{Mod.
  Phys. Lett. A} {\bfseries 36} (2021) 2130006}
  [\href{https://arxiv.org/abs/2012.05735}{{\ttfamily 2012.05735}}].

\bibitem{Darme:2021gtt}
L.~Darm{\'e}, B.~Fuks and F.~Maltoni, \emph{{Top-philic heavy resonances in
  four-top final states and their EFT interpretation}},
  \href{https://doi.org/10.1007/JHEP09(2021)143}{\emph{JHEP} {\bfseries 09}
  (2021) 143} [\href{https://arxiv.org/abs/2104.09512}{{\ttfamily
  2104.09512}}].

\bibitem{Cao:2021qqt}
Q.-H.~Cao, J.-N.~Fu, Y.~Liu, X.-H.~Wang and R.~Zhang, \emph{{Probing top-philic
  new physics via four-top-quark production}},
  \href{https://doi.org/10.1088/1674-1137/ac0c6f}{\emph{Chin. Phys. C}
  {\bfseries 45} (2021) 093107}
  [\href{https://arxiv.org/abs/2105.03372}{{\ttfamily 2105.03372}}].

\bibitem{Carpenter:2021vga}
L.M.~Carpenter, T.~Murphy and M.J.~Smylie, \emph{{$ t\overline{t}t\overline{t}
  $ signatures through the lens of color-octet scalars}},
  \href{https://doi.org/10.1007/JHEP01(2022)047}{\emph{JHEP} {\bfseries 01}
  (2022) 047} [\href{https://arxiv.org/abs/2107.13565}{{\ttfamily
  2107.13565}}].

\bibitem{Aoude:2022deh}
R.~Aoude, H.~El~Faham, F.~Maltoni and E.~Vryonidou, \emph{{Complete SMEFT
  predictions for four top quark production at hadron colliders}},
  \href{https://doi.org/10.1007/JHEP10(2022)163}{\emph{JHEP} {\bfseries 10}
  (2022) 163} [\href{https://arxiv.org/abs/2208.04962}{{\ttfamily
  2208.04962}}].

\bibitem{Bally:2022naz}
A.~Bally, Y.~Chung and F.~Goertz, \emph{{Hierarchy problem and the top Yukawa
  coupling: An alternative to top partner solutions}},
  \href{https://doi.org/10.1103/PhysRevD.108.055008}{\emph{Phys. Rev. D}
  {\bfseries 108} (2023) 055008}
  [\href{https://arxiv.org/abs/2211.17254}{{\ttfamily 2211.17254}}].

\bibitem{Aleshko:2023rkv}
A.~Aleshko, E.~Boos, V.~Bunichev and L.~Dudko, \emph{{Prospects for
  establishing limits on the SMEFT operators from the production processes of
  three and four top quarks in hadron collisions}},
  \href{https://doi.org/10.1142/S0217751X24501197}{\emph{Int. J. Mod. Phys. A}
  {\bfseries 39} (2024) 2450119}
  [\href{https://arxiv.org/abs/2309.12514}{{\ttfamily 2309.12514}}].

\bibitem{Choudhury:2024mox}
D.~Choudhury, K.~Deka and L.K.~Saini, \emph{{Boosted four-top production at the
  LHC: A window to Randall-Sundrum or extended color symmetry}},
  \href{https://doi.org/10.1103/PhysRevD.110.075020}{\emph{Phys. Rev. D}
  {\bfseries 110} (2024) 075020}
  [\href{https://arxiv.org/abs/2404.04409}{{\ttfamily 2404.04409}}].

\bibitem{Degrande:2024mbg}
C.~Degrande, R.~Rosenfeld and A.~Vasquez, \emph{{Collider sensitivity to SMEFT
  heavy-quark operators at one-loop in top-quark processes}},
  \href{https://doi.org/10.1007/JHEP07(2024)114}{\emph{JHEP} {\bfseries 07}
  (2024) 114} [\href{https://arxiv.org/abs/2402.06528}{{\ttfamily
  2402.06528}}].

\bibitem{DiNoi:2025uhu}
S.~Di~Noi, H.~El~Faham, R.~Gr{\"o}ber, M.~Vitti and E.~Vryonidou,
  \emph{{Constraining four-heavy-quark operators with top-quark, Higgs, and
  electroweak precision data}},
  \href{https://doi.org/10.1007/JHEP01(2026)025}{\emph{JHEP} {\bfseries 01}
  (2026) 025} [\href{https://arxiv.org/abs/2507.01137}{{\ttfamily
  2507.01137}}].

\bibitem{Darme:2025leu}
L.~Darm{\'e}, B.~Fuks, H.-L.~Li, M.~Maltoni and J.~Touch{\`e}que,
  \emph{{Searching for top-philic heavy resonances in boosted four-top final
  states}}, \href{https://doi.org/10.1007/JHEP11(2025)091}{\emph{JHEP}
  {\bfseries 11} (2025) 091}
  [\href{https://arxiv.org/abs/2507.05334}{{\ttfamily 2507.05334}}].

\bibitem{Blekman:2022jag}
F.~Blekman, F.~D{\'e}liot, V.~Dutta and E.~Usai, \emph{{Four-top quark physics
  at the LHC}}, \href{https://doi.org/10.3390/universe8120638}{\emph{Universe}
  {\bfseries 8} (2022) 638} [\href{https://arxiv.org/abs/2208.04085}{{\ttfamily
  2208.04085}}].

\bibitem{Bevilacqua:2012em}
G.~Bevilacqua and M.~Worek, \emph{{Constraining BSM Physics at the LHC: Four
  top final states with NLO accuracy in perturbative QCD}},
  \href{https://doi.org/10.1007/JHEP07(2012)111}{\emph{JHEP} {\bfseries 07}
  (2012) 111} [\href{https://arxiv.org/abs/1206.3064}{{\ttfamily 1206.3064}}].

\bibitem{Maltoni:2015ena}
F.~Maltoni, D.~Pagani and I.~Tsinikos, \emph{{Associated production of a
  top-quark pair with vector bosons at NLO in QCD: impact on $
  \mathrm{t}\overline{\mathrm{t}}\mathrm{H} $ searches at the LHC}},
  \href{https://doi.org/10.1007/JHEP02(2016)113}{\emph{JHEP} {\bfseries 02}
  (2016) 113} [\href{https://arxiv.org/abs/1507.05640}{{\ttfamily
  1507.05640}}].

\bibitem{Frederix:2017wme}
R.~Frederix, D.~Pagani and M.~Zaro, \emph{{Large NLO corrections in
  $t\bar{t}W^{\pm}$ and $t\bar{t}t\bar{t}$ hadroproduction from supposedly
  subleading EW contributions}},
  \href{https://doi.org/10.1007/JHEP02(2018)031}{\emph{JHEP} {\bfseries 02}
  (2018) 031} [\href{https://arxiv.org/abs/1711.02116}{{\ttfamily
  1711.02116}}].

\bibitem{vanBeekveld:2022hty}
M.~van Beekveld, A.~Kulesza and L.M.~Valero, \emph{{Threshold Resummation for
  the Production of Four Top Quarks at the LHC}},
  \href{https://doi.org/10.1103/PhysRevLett.131.211901}{\emph{Phys. Rev. Lett.}
  {\bfseries 131} (2023) 211901}
  [\href{https://arxiv.org/abs/2212.03259}{{\ttfamily 2212.03259}}].

\bibitem{vanBeekveld:2025ghw}
M.~van Beekveld, A.~Kulesza, M.~Lupattelli and T.~Saracco,
  \emph{{Invariant-mass threshold resummation for the production of four top
  quarks at the LHC}},
  \href{https://doi.org/10.1007/JHEP10(2025)209}{\emph{JHEP} {\bfseries 10}
  (2025) 209} [\href{https://arxiv.org/abs/2505.10381}{{\ttfamily
  2505.10381}}].

\bibitem{Alsairafi:2025rjd}
M.~Alsairafi, N.~Dimitrakopoulos and M.~Worek, \emph{{Modelling top-quark
  decays in $t\bar{t}t\bar{t}$ production at the LHC}},
  \href{https://arxiv.org/abs/2507.04849}{{\ttfamily 2507.04849}}.

\bibitem{ATLAS:2023ajo}
{\scshape ATLAS} collaboration, \emph{{Observation of four-top-quark production
  in the multilepton final state with the ATLAS detector}},
  \href{https://doi.org/10.1140/epjc/s10052-023-11573-0}{\emph{Eur. Phys. J. C}
  {\bfseries 83} (2023) 496}
  [\href{https://arxiv.org/abs/2303.15061}{{\ttfamily 2303.15061}}].

\bibitem{CMS:2023ftu}
{\scshape CMS} collaboration, \emph{{Observation of four top quark production
  in proton-proton collisions at $\sqrt{s}=13\,$TeV}},
  \href{https://doi.org/10.1016/j.physletb.2023.138290}{\emph{Phys. Lett. B}
  {\bfseries 847} (2023) 138290}
  [\href{https://arxiv.org/abs/2305.13439}{{\ttfamily 2305.13439}}].

\bibitem{CMS:2025rug}
{\scshape CMS} collaboration, \emph{{Search for physics beyond the standard
  model in four and three top quark production events using proton-proton
  collisions at $\sqrt{s}=13\,\mathrm{TeV}$}}, {\emph{CMS-PAS-TOP-24-008}
  (2025) }.

\bibitem{Frixione:2008yi}
S.~Frixione, E.~Laenen, P.~Motylinski, B.R.~Webber and C.D.~White,
  \emph{{Single-top hadroproduction in association with a W boson}},
  \href{https://doi.org/10.1088/1126-6708/2008/07/029}{\emph{JHEP} {\bfseries
  07} (2008) 029} [\href{https://arxiv.org/abs/0805.3067}{{\ttfamily
  0805.3067}}].

\bibitem{White:2009yt}
C.D.~White, S.~Frixione, E.~Laenen and F.~Maltoni, \emph{{Isolating Wt
  production at the LHC}},
  \href{https://doi.org/10.1088/1126-6708/2009/11/074}{\emph{JHEP} {\bfseries
  11} (2009) 074} [\href{https://arxiv.org/abs/0908.0631}{{\ttfamily
  0908.0631}}].

\bibitem{Re:2010bp}
E.~Re, \emph{{Single-top Wt-channel production matched with parton showers
  using the POWHEG method}},
  \href{https://doi.org/10.1140/epjc/s10052-011-1547-z}{\emph{Eur. Phys. J. C}
  {\bfseries 71} (2011) 1547}
  [\href{https://arxiv.org/abs/1009.2450}{{\ttfamily 1009.2450}}].

\bibitem{Hollik:2012rc}
W.~Hollik, J.M.~Lindert and D.~Pagani, \emph{{NLO corrections to squark-squark
  production and decay at the LHC}},
  \href{https://doi.org/10.1007/JHEP03(2013)139}{\emph{JHEP} {\bfseries 03}
  (2013) 139} [\href{https://arxiv.org/abs/1207.1071}{{\ttfamily 1207.1071}}].

\bibitem{Frixione:2019fxg}
S.~Frixione, B.~Fuks, V.~Hirschi, K.~Mawatari, H.-S.~Shao, P.A.~Sunder et~al.,
  \emph{{Automated simulations beyond the Standard Model: supersymmetry}},
  \href{https://doi.org/10.1007/JHEP12(2019)008}{\emph{JHEP} {\bfseries 12}
  (2019) 008} [\href{https://arxiv.org/abs/1907.04898}{{\ttfamily
  1907.04898}}].

\bibitem{Faham:2021zet}
H.E.~Faham, F.~Maltoni, K.~Mimasu and M.~Zaro, \emph{{Single top production in
  association with a WZ pair at the LHC in the SMEFT}},
  \href{https://doi.org/10.1007/JHEP01(2022)100}{\emph{JHEP} {\bfseries 01}
  (2022) 100} [\href{https://arxiv.org/abs/2111.03080}{{\ttfamily
  2111.03080}}].

\bibitem{Barger:2010uw}
V.~Barger, W.-Y.~Keung and B.~Yencho, \emph{{Triple-Top Signal of New Physics
  at the LHC}},
  \href{https://doi.org/10.1016/j.physletb.2010.03.001}{\emph{Phys. Lett. B}
  {\bfseries 687} (2010) 70} [\href{https://arxiv.org/abs/1001.0221}{{\ttfamily
  1001.0221}}].

\bibitem{Boos:2021yat}
E.~Boos and L.~Dudko, \emph{{Triple top quark production in standard model}},
  \href{https://doi.org/10.1142/S0217751X22500233}{\emph{Int. J. Mod. Phys. A}
  {\bfseries 37} (2022) 2250023}
  [\href{https://arxiv.org/abs/2107.07629}{{\ttfamily 2107.07629}}].

\bibitem{Demartin:2016axk}
F.~Demartin, B.~Maier, F.~Maltoni, K.~Mawatari and M.~Zaro, \emph{{tWH
  associated production at the LHC}},
  \href{https://doi.org/10.1140/epjc/s10052-017-4601-7}{\emph{Eur. Phys. J. C}
  {\bfseries 77} (2017) 34} [\href{https://arxiv.org/abs/1607.05862}{{\ttfamily
  1607.05862}}].

\bibitem{madstr}
M.~Zaro, \emph{{MadSTR}}, {\emph{\url{https://github.com/mg5amcnlo/MadSTR}} }.

\bibitem{Alwall:2014hca}
J.~Alwall, R.~Frederix, S.~Frixione, V.~Hirschi, F.~Maltoni, O.~Mattelaer
  et~al., \emph{{The automated computation of tree-level and next-to-leading
  order differential cross sections, and their matching to parton shower
  simulations}}, \href{https://doi.org/10.1007/JHEP07(2014)079}{\emph{JHEP}
  {\bfseries 07} (2014) 079} [\href{https://arxiv.org/abs/1405.0301}{{\ttfamily
  1405.0301}}].

\bibitem{Frederix:2018nkq}
R.~Frederix, S.~Frixione, V.~Hirschi, D.~Pagani, H.S.~Shao and M.~Zaro,
  \emph{{The automation of next-to-leading order electroweak calculations}},
  \href{https://doi.org/10.1007/JHEP11(2021)085}{\emph{JHEP} {\bfseries 07}
  (2018) 185} [\href{https://arxiv.org/abs/1804.10017}{{\ttfamily
  1804.10017}}].

\bibitem{Ball:2017nwa}
{\scshape NNPDF} collaboration, \emph{{Parton distributions from high-precision
  collider data}},
  \href{https://doi.org/10.1140/epjc/s10052-017-5199-5}{\emph{Eur. Phys. J. C}
  {\bfseries 77} (2017) 663}
  [\href{https://arxiv.org/abs/1706.00428}{{\ttfamily 1706.00428}}].

\bibitem{Manohar:2016nzj}
A.~Manohar, P.~Nason, G.P.~Salam and G.~Zanderighi, \emph{{How bright is the
  proton? A precise determination of the photon parton distribution function}},
  \href{https://doi.org/10.1103/PhysRevLett.117.242002}{\emph{Phys. Rev. Lett.}
  {\bfseries 117} (2016) 242002}
  [\href{https://arxiv.org/abs/1607.04266}{{\ttfamily 1607.04266}}].

\bibitem{Manohar:2017eqh}
A.V.~Manohar, P.~Nason, G.P.~Salam and G.~Zanderighi, \emph{{The Photon Content
  of the Proton}}, \href{https://doi.org/10.1007/JHEP12(2017)046}{\emph{JHEP}
  {\bfseries 12} (2017) 046}
  [\href{https://arxiv.org/abs/1708.01256}{{\ttfamily 1708.01256}}].

\bibitem{Buckley:2014ana}
A.~Buckley, J.~Ferrando, S.~Lloyd, K.~Nordstr{\"o}m, B.~Page, M.~R{\"u}fenacht
  et~al., \emph{{LHAPDF6: parton density access in the LHC precision era}},
  \href{https://doi.org/10.1140/epjc/s10052-015-3318-8}{\emph{Eur. Phys. J. C}
  {\bfseries 75} (2015) 132} [\href{https://arxiv.org/abs/1412.7420}{{\ttfamily
  1412.7420}}].

\bibitem{Frixione:2014qaa}
S.~Frixione, V.~Hirschi, D.~Pagani, H.S.~Shao and M.~Zaro, \emph{{Weak
  corrections to Higgs hadroproduction in association with a top-quark pair}},
  \href{https://doi.org/10.1007/JHEP09(2014)065}{\emph{JHEP} {\bfseries 09}
  (2014) 065} [\href{https://arxiv.org/abs/1407.0823}{{\ttfamily 1407.0823}}].

\bibitem{Frixione:2015zaa}
S.~Frixione, V.~Hirschi, D.~Pagani, H.S.~Shao and M.~Zaro, \emph{{Electroweak
  and QCD corrections to top-pair hadroproduction in association with heavy
  bosons}}, \href{https://doi.org/10.1007/JHEP06(2015)184}{\emph{JHEP}
  {\bfseries 06} (2015) 184}
  [\href{https://arxiv.org/abs/1504.03446}{{\ttfamily 1504.03446}}].

\bibitem{Pagani:2016caq}
D.~Pagani, I.~Tsinikos and M.~Zaro, \emph{{The impact of the photon PDF and
  electroweak corrections on $t \bar{t}$ distributions}},
  \href{https://doi.org/10.1140/epjc/s10052-016-4318-z}{\emph{Eur. Phys. J. C}
  {\bfseries 76} (2016) 479}
  [\href{https://arxiv.org/abs/1606.01915}{{\ttfamily 1606.01915}}].

\bibitem{Frederix:2016ost}
R.~Frederix, S.~Frixione, V.~Hirschi, D.~Pagani, H.-S.~Shao and M.~Zaro,
  \emph{{The complete NLO corrections to dijet hadroproduction}},
  \href{https://doi.org/10.1007/JHEP04(2017)076}{\emph{JHEP} {\bfseries 04}
  (2017) 076} [\href{https://arxiv.org/abs/1612.06548}{{\ttfamily
  1612.06548}}].

\bibitem{Czakon:2017wor}
M.~Czakon, D.~Heymes, A.~Mitov, D.~Pagani, I.~Tsinikos and M.~Zaro,
  \emph{{Top-pair production at the LHC through NNLO QCD and NLO EW}},
  \href{https://doi.org/10.1007/JHEP10(2017)186}{\emph{JHEP} {\bfseries 10}
  (2017) 186} [\href{https://arxiv.org/abs/1705.04105}{{\ttfamily
  1705.04105}}].

\bibitem{Broggio:2019ewu}
A.~Broggio, A.~Ferroglia, R.~Frederix, D.~Pagani, B.D.~Pecjak and I.~Tsinikos,
  \emph{{Top-quark pair hadroproduction in association with a heavy boson at
  NLO+NNLL including EW corrections}},
  \href{https://doi.org/10.1007/JHEP08(2019)039}{\emph{JHEP} {\bfseries 08}
  (2019) 039} [\href{https://arxiv.org/abs/1907.04343}{{\ttfamily
  1907.04343}}].

\bibitem{Frederix:2019ubd}
R.~Frederix, D.~Pagani and I.~Tsinikos, \emph{{Precise predictions for
  single-top production: the impact of EW corrections and QCD shower on the
  $t$-channel signature}},
  \href{https://doi.org/10.1007/JHEP09(2019)122}{\emph{JHEP} {\bfseries 09}
  (2019) 122} [\href{https://arxiv.org/abs/1907.12586}{{\ttfamily
  1907.12586}}].

\bibitem{Pagani:2020rsg}
D.~Pagani, H.-S.~Shao and M.~Zaro, \emph{{RIP $ Hb\overline{b} $: how other
  Higgs production modes conspire to kill a rare signal at the LHC}},
  \href{https://doi.org/10.1007/JHEP11(2020)036}{\emph{JHEP} {\bfseries 11}
  (2020) 036} [\href{https://arxiv.org/abs/2005.10277}{{\ttfamily
  2005.10277}}].

\bibitem{Pagani:2020mov}
D.~Pagani, I.~Tsinikos and E.~Vryonidou, \emph{{NLO QCD+EW predictions for
  $tHj$ and $tZj$ production at the LHC}},
  \href{https://doi.org/10.1007/JHEP08(2020)082}{\emph{JHEP} {\bfseries 08}
  (2020) 082} [\href{https://arxiv.org/abs/2006.10086}{{\ttfamily
  2006.10086}}].

\bibitem{Pagani:2021iwa}
D.~Pagani, H.-S.~Shao, I.~Tsinikos and M.~Zaro, \emph{{Automated EW corrections
  with isolated photons: t$ \overline{t} ${\ensuremath{\gamma}}, t$
  \overline{t} ${\ensuremath{\gamma}}{\ensuremath{\gamma}} and
  t{\ensuremath{\gamma}}j as case studies}},
  \href{https://doi.org/10.1007/JHEP09(2021)155}{\emph{JHEP} {\bfseries 09}
  (2021) 155} [\href{https://arxiv.org/abs/2106.02059}{{\ttfamily
  2106.02059}}].

\bibitem{ElFaham:2024egs}
H.~El~Faham, K.~Mimasu, D.~Pagani, C.~Severi, E.~Vryonidou and M.~Zaro,
  \emph{{Electroweak corrections in the SMEFT: four-fermion operators at high
  energies}}, \href{https://doi.org/10.1007/JHEP06(2025)241}{\emph{JHEP}
  {\bfseries 06} (2025) 241}
  [\href{https://arxiv.org/abs/2412.16076}{{\ttfamily 2412.16076}}].

\bibitem{Frixione:1992pj}
S.~Frixione, P.~Nason and G.~Ridolfi, \emph{{Strong corrections to W Z
  production at hadron colliders}},
  \href{https://doi.org/10.1016/0550-3213(92)90668-2}{\emph{Nucl. Phys. B}
  {\bfseries 383} (1992) 3}.

\bibitem{Rubin:2010xp}
M.~Rubin, G.P.~Salam and S.~Sapeta, \emph{{Giant QCD $K$-factors beyond NLO}},
  \href{https://doi.org/10.1007/JHEP09(2010)084}{\emph{JHEP} {\bfseries 09}
  (2010) 084} [\href{https://arxiv.org/abs/1006.2144}{{\ttfamily 1006.2144}}].

\bibitem{Durupt:2025wuk}
V.~Durupt, F.~Maltoni and O.~Mattelaer, \emph{{Automated computation of
  spin-density matrices and quantum observables for collider physics}},
  \href{https://doi.org/10.1007/JHEP04(2026)103}{\emph{JHEP} {\bfseries 04}
  (2026) 103} [\href{https://arxiv.org/abs/2510.17730}{{\ttfamily
  2510.17730}}].

\end{thebibliography}\endgroup
\end{document}